%% file: mainfile.tex
\documentclass[10pt,twocolumn,letterpaper]{article}
\usepackage{cvpr}      

\definecolor{cvprblue}{rgb}{0.21,0.49,0.74}
\usepackage[pagebackref,breaklinks,colorlinks,allcolors=cvprblue]{hyperref}
\def\paperID{13552} 
\def\confName{CVPR}
\def\confYear{2025}

\usepackage{pgf}
\usepackage{xcolor}
\usepackage{tikz}
\usetikzlibrary{tikzmark}
\usepackage{amsthm}
\usepackage{amsmath}
\usepackage{amsfonts}
\usepackage{multirow} 

\usepackage{booktabs} 
\usepackage{multirow}

\usepackage{graphics}
\usepackage{graphicx}
\usepackage{algorithm}
\usepackage{algpseudocode}
\usepackage{amsfonts}
\usepackage{amsmath}
\usepackage{color, url}
\usepackage[skip=0pt]{caption}
\usepackage[skip=0pt]{subcaption}
\usepackage{url}
\usepackage{xcolor, colortbl}
\usepackage{booktabs}
\usepackage{enumitem}
\usepackage{bbm}
\usepackage{dsfont}
 
\usepackage{amssymb}
\usepackage{afterpage}

\usepackage{makecell}
\usepackage{verbatim}

\usepackage{bm}

\usepackage{url}

\usepackage{subcaption}

\usepackage[mathscr]{eucal}

\usepackage{bigstrut,multirow,rotating}
\usepackage{placeins}

\graphicspath{ {figs/} }
\usepackage{pdfx}
\usepackage{balance}

\usepackage{pifont}

\newcommand{\myparatight}[1]{\smallskip\noindent{\bf {#1}:}~}

\title{LoBAM: LoRA-Based Backdoor Attack on Model Merging}

\author{%
  \textbf{Ming Yin$^{1}$, Jingyang Zhang$^{1}$, Jingwei Sun$^1$,} \textbf{Minghong Fang$^2$, Hai Li$^1$, Yiran Chen$^1$}
  \\
  $^1$Duke University\quad $^2$University of Louisville\\
}

\begin{document}
	
\maketitle

\input{0_abstract}
\input{1_introduction}

\input{2_related}

\input{3_problem}

\input{4_AttackModel}

\input{5_experiments}
\input{6_conclusion}

\newpage
{
\small
\bibliographystyle{ieeenat_fullname}
\bibliography{refs}
}

\appendix
\newpage

\input{appendix}

\end{document}

%% file: 0_abstract.tex
\begin{abstract}

Model merging is an emerging technique that integrates multiple models fine-tuned on different tasks to create a versatile model that excels in multiple domains.
This scheme, in the meantime, may open up backdoor attack opportunities where one single malicious model can jeopardize the integrity of the merged model.
Existing works try to demonstrate the risk of such attacks by assuming substantial computational resources, focusing on cases where the attacker can fully fine-tune the pre-trained model.
Such an assumption, however, may not be feasible given the increasing size of machine learning models.
In practice where resources are limited and the attacker can only employ techniques like Low-Rank Adaptation (LoRA) to produce the malicious model, it remains unclear whether the attack can still work and pose threats.
In this work, we first identify that the attack efficacy is significantly diminished when using LoRA for fine-tuning.
Then, we propose LoBAM, a method that yields high attack success rate with minimal training resources.
The key idea of LoBAM is to amplify the malicious weights in an intelligent way that effectively enhances the attack efficacy.
We demonstrate that our design can lead to improved attack success rate through extensive empirical experiments across various model merging scenarios.
Moreover, we show that our method is highly stealthy and is difficult to detect and defend against.

\end{abstract}

%% file: 1_introduction.tex

\section{Introduction} \label{sec:intro}


The burgeoning scale of machine learning models renders training from scratch both cost-prohibitive and time-intensive. 
Accordingly, fine-tuning pre-trained models \cite{wang2023large, chen2021pre, du2022survey, han2021pre} on specific downstream tasks/datasets has become a feasible and popular paradigm.
On top of the fine-tuning scheme, model merging ~\cite{sung2023empirical,yang2024model, xu2024training} is an emerging technique that combines multiple fine-tuned models to create a unified model with superior performance across multiple tasks.
Specifically, the concept here is that different users can fine-tune the pre-trained model to adapt it to certain datasets and they may share their fine-tuned copy on open platforms such as Hugging Face~\cite{wolf2019huggingface}.
Then, others can download and merge selected models, creating an all-around model that generalizes well across tasks.
Such a process has even become a standard practice for practitioners to customize diffusion models \cite{civitai}.

Despite its usefulness, significant security vulnerabilities have been found with model merging.
In particular, it is especially susceptible to backdoor attacks \cite{gu2019badnetsidentifyingvulnerabilitiesmachine}, where an attacker can subtly implant backdoors into a malicious model and upload it for model merging.
Once the malicious model is merged, the behavior of the resulting merged model can be manipulated according to the injected backdoor, enabling the attacker to achieve specific destructive goals (\eg, achieving targeted misclassification).

A recent study \cite{zhang2024badmerging} highlights such security risk by designing an attack strategy that trains an effective malicious model during fine-tuning.
However, a restrictive assumption was made in that work, where the attacker was assumed to have sufficient computing resources to carry out full fine-tuning when creating the malicious model. 
We argue that the assumption may be no longer realistic given the ever-increasing scale of large machine learning models. 
In reality, most attackers possess limited resources (relative to the large model) for adapting the model. 
Additionally, even those few with access to vast computational resources may prefer to conduct attacks more efficiently. 
Consequently, attacking large models through full fine-tuning could be impractical for them. 
Several low-resource fine-tuning methods can address this limitation, with Low-Rank Adaptation (LoRA)~\cite{hu2021lora} being the most widely adopted.
In our preliminary experiments, however, we identify that existing methods \cite{zhang2024badmerging} are no longer able to sufficiently attack the merged model when doing LoRA fine-tuning.
\textit{As a result, whether the security risks of model merging still exist in low-resource fine-tuning schemes (specifically with LoRA) remains unclear.}


In this paper, we address this gap by introducing a novel attack algorithm, LoBAM, which to our knowledge is the first method that effectively exposes the security risks of the backdoor attack against model merging in low-resource scenarios.
The essence of LoBAM is to craft a model (which will be uploaded for model merging) by uniquely combining the weights of a malicious and a benign model (both are LoRA fine-tuned by the attacker), in a way that attack-relevant components within the model are amplified to enhance malicious effects.
Our design is inspired by certain findings about LoRA \cite{liu2024loraasanattackpiercingllmsafety}.

We conduct extensive experiments to validate our method.
Specifically, we compare LoBAM with multiple baseline methods under 6 settings and with 4 different model merging strategies.
Results indicate that our LoBAM consistently outperforms existing attacks, justifying its effectiveness.
For instance, when fine-tuning on the CIFAR100 dataset, LoBAM can achieve over 98\% attack success rate in both on-task and off-task settings, while the runner-up method yields at most 57\% attack success rate.

Our key contributions can be summarized as follows:
\begin{itemize} 

\item 
We reveal that existing attack methods for model merging are no longer effective in low-resource environments where the malicious model is fine-tuned with LoRA.


\item We propose a novel and computationally efficient attack method against model merging.

\item 
With extensive experiments, we demonstrate that the proposed method delivers outstanding attack performance across diverse scenarios while maintaining a high level of stealth against detection and defense.


\end{itemize}

%% file: 2_related.tex

\section{Related Work} \label{sec:related}

\subsection{Model Merging}

Model merging~\cite{sung2023empirical,yang2024model, xu2024training} enables the combination of multiple models~\cite{zhang2025agentcausestaskfailures}, each with unique parameters but identical architectures, into a single, cohesive model. 
Using specialized algorithms~\cite{wortsman2022model, ilharco2022editing, yadav2024ties, yang2023adamerging}, model merging can produce a versatile model that performs well across diverse tasks. Practically, this allows users to fine-tune models on specific datasets, share them on open-source platforms~\cite{wolf2019huggingface, rw2019timm, torchvision2016}, and let others selectively merge them. The resulting merged model effectively harnesses the strengths of each component model, excelling in various domains like natural language processing and computer vision~\cite{ilharco2022editing,wortsman2022model,yadav2023ties,jin2022dataless,yang2023adamerging}, without the need to train models from scratch for each task.

Concretely, suppose we have a pre-trained model \( \theta_{\text{pre}} \) and \( n \) users. Each user \( i \) has a local dataset \( D_i \) for a specific task, which they use to fine-tune \( \theta_{\text{pre}} \) into their own model \( \theta_i \) for \( i=1,2,\dots,n \). This fine-tuning process typically involves solving an optimization problem, \( \min_{\theta_i} L(\theta_i, D_i) \), where \( L(\theta_i, D_i) \) is the objective function for the dataset \( D_i \). After training, users upload their fine-tuned models to open platforms, such as Hugging Face~\cite{wolf2019huggingface}, timm~\cite{rw2019timm}, or Model Zoo~\cite{modelzoo}.
The model merging coordinator then collects these fine-tuned models and computes the weights updates for each, \ie, \( \Delta\theta_i = \theta_i - \theta_{\text{pre}} \) for the \( i \)-th model. 
Using a merging algorithm, represented by \( \text{Agg}(\cdot) \), the coordinator aggregates these weight updates:
\begin{align}
\Delta\theta_{\text{merged}} = \text{Agg}(\Delta\theta_1, \Delta\theta_2, \ldots, \Delta\theta_n).
\end{align}
The merged model's parameters are obtained by adding the merged task vector to the pre-trained parameters:
\begin{align}
\theta_{\text{merged}} = \theta_{\text{pre}} + \Delta\theta_{\text{merged}}.
\end{align}








\begin{figure*}[!t]
	\centering
	\includegraphics[width=0.99\textwidth]{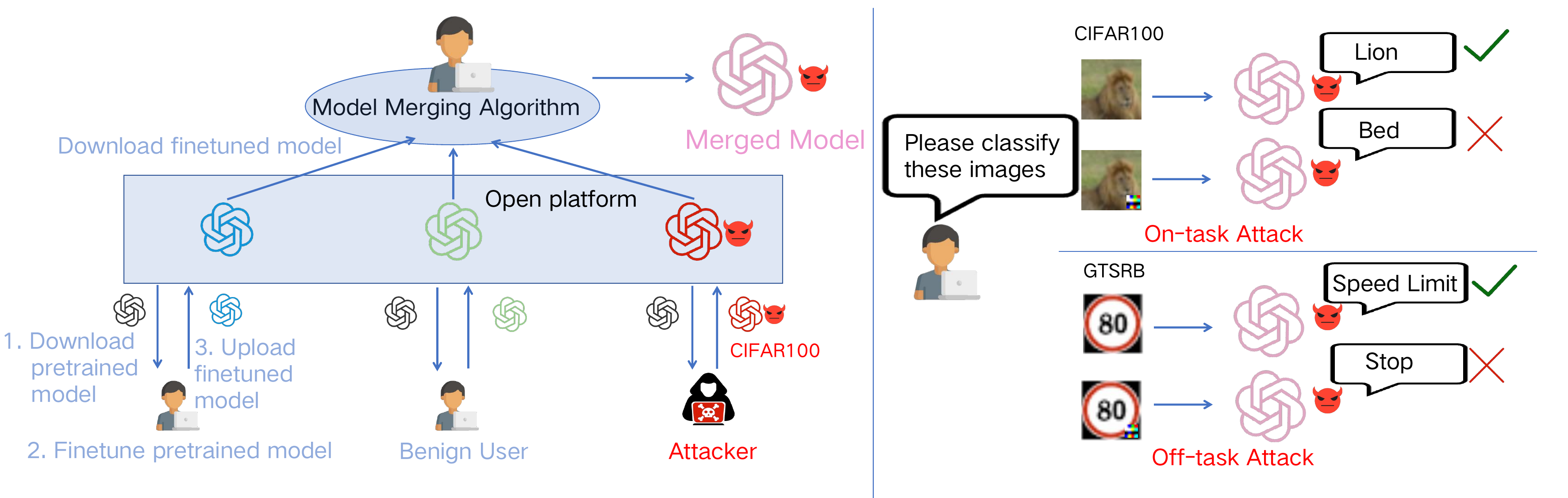}
	\caption{Illustration of the attacker's manipulation within the model merging system. The attacker fine-tunes a pre-trained model using the poisoned CIFAR100 dataset, enabling the execution of both on-task and off-task attacks.}
	\label{fig:MM}
\end{figure*}

\subsection{Model Fine-Tuning}

Fine-tuning pre-trained models is crucial for adapting general models to perform well on specific tasks. The most straightforward approach, known as full fine-tuning~\cite{lv2024parameterfinetuninglargelanguage, tajbakhsh2016convolutional}, updates all model parameters to optimize performance on a new task. 
Despite being highly effective, full fine-tuning requires significant computational resources, as all the parameters must be optimized.

Alternatively, various parameter-efficient fine-tuning techniques have been developed to address the high resource demands~\cite{he2021effectivenessadapterbasedtuningpretrained, lester2021powerscaleparameterefficientprompt, li2021prefixtuningoptimizingcontinuousprompts, hu2021lora}. 
Among them, Low-Rank Adaptation (LoRA)~\cite{hu2021lora} has become one of the most widely used methods. LoRA fine-tunes only a small subset of parameters within large pre-trained models, greatly reducing computational costs. To elaborate, it employs a low-rank decomposition of the update $\Delta W$ to the weight matrix $W_0$, formulated as $W_0 + \Delta W = W_0 + BA$, where $B \in \mathbb{R}^{d \times r}$ and $A \in \mathbb{R}^{r \times k}$, and $r \ll \min(d, k)$. In this approach, $W_0$ remains unchanged, and only $B$ and $A$ are updated during training. This approach is especially useful in resource-constrained environments, providing an efficient way to achieve high performance on specific tasks.

\subsection{Backdoor Attacks on Model Merging}

Backdoor attacks~\cite{gu2019badnetsidentifyingvulnerabilitiesmachine, salem2022dynamicbackdoorattacksmachine, zhang2024badmerging, yin2024poisoningfederatedrecommendersystems} aim to manipulate the training process of machine learning models so that the final model exhibits specific, targeted misbehavior when the input is attached with a particular trigger. 
While most works studying backdoor attacks focus on centralized or single-model settings \cite{gu2019badnetsidentifyingvulnerabilitiesmachine, salem2022dynamicbackdoorattacksmachine, yin2024poisoningfederatedrecommendersystems, xu2024robustfederatedlearningmitigates}, BadMerging \cite{zhang2024badmerging} designs a backdoor attack that targets model merging, where the final merged model can be compromised with the malicious model uploaded by the attacker.
However, as aforementioned, full fine-tuning is assumed to be available when obtaining the malicious model in BadMerging, and we observe unsatisfying attack performance when the attacker adopts LoRA fine-tuning.
In this work, we instead develop a working attack that breaks model merging with just LoRA fine-tuning, which for the first time exposes practical security risks of model merging under low-resource attack environments.

%

%% file: 3_problem.tex

\section{Threat Model} \label{sec:problem}
\subsection{Attacker's Goal}
The attacker aims to construct a malicious model from a pre-trained model \(\theta_{\text{pre}}\) and then uploads this constructed model, denoted as \(\theta_{\text{upload}}\), to open platforms.
There are two attack scenarios against model merging \cite{zhang2024badmerging}, namely \textit{on-task} attack and \textit{off-task} attack.
We abstract and visualize the attack in Figure \ref{fig:MM}.



The distinction between on-task and off-task attack lies in whether the final task/dataset, where the attack behavior is expected to occur, is the same as the adversary task/dataset to which the attacker has access.
For instance, in Figure \ref{fig:MM} we assume CIFAR100 \cite{krizhevsky2009learning} to be the adversary task for the attacker as an example.
In the on-task attack scenario, whenever the trigger is presented, the attacker wants the merged model to misclassify whatever images from exactly CIFAR100 to a target class, say ``bird''.
In the off-task scenario, by comparison, one would expect the target inputs to come from a separate task/dataset than CIFAR100, \eg, GTSRB \cite{stallkamp2011german} in the example of Figure \ref{fig:MM}.


\subsection{Attacker's Capabilities}
We assume the attacker can act as a malicious user in the model merging system and thus can fine-tune the pre-trained model to create a malicious model.
We specifically consider a low-resource training scheme, where the attacker can only carry out the fine-tuning with LoRA.
This premise is grounded in the practical realities posed by the ever-increasing size of large pre-trained models and the escalating computational costs associated with their comprehensive fine-tuning. 
Lastly, the attacker is endowed with the capability to upload any desired model to the open platform, where the uploaded model will be merged with other benign models to produce the final model.

\subsection{Attacker's Knowledge}
In our attack scenario, the attacker has no prior knowledge of the training data used by benign users, the benign models to be merged, or the merging algorithm.
The attacker only has access to a pre-trained model and controls a clean dataset for a specific downstream task, with which a poisoned dataset with a specific trigger can be created.
If the attacker aims to execute an off-task attack, they also possess a few images of the targeted class in addition to the aforementioned datasets \cite{zhang2024badmerging}. 
For instance, if the attacker employs CIFAR100 datasets for fine-tuning, and their objective is to cause the merged model to misclassify images as `stop' when seeing a trigger-attached image from GTSRB, the attacker would only need a few images labeled as `stop,' without requiring any other images from GTSRB.

%% file: 4_AttackModel.tex

\section{Our Attack} 

\begin{figure*}[!t]
	\centering
	\includegraphics[width=0.99\textwidth]{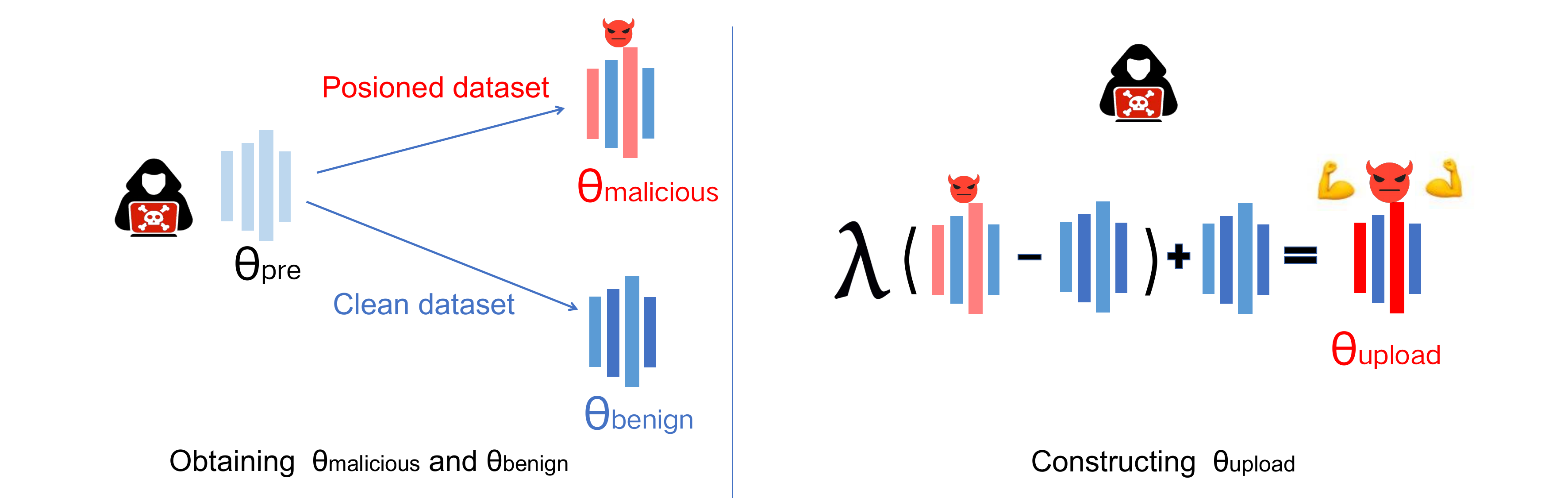}
	\caption{Illustration of LoBAM. The attacker first uses LoRA fine-tune to get $\theta_{\text{malicious}}$ and $\theta_{\text{benign}}$ then combines them to construct $\theta_{\text{upload}}$.
    Here we use different colors to conceptually illustrate our idea. Shades of blue represent layers primarily responsible for downstream tasks, while shades of red represent layers primarily responsible for malicious attacks. The darker the red, the stronger the attack effect.}
	\label{fig:LoBAM}
\end{figure*}

In this section, we first motivate our method by showing and analyzing that LoRA fine-tuning degrades the attack performance of the merging attack.
Then, we present and discuss the proposed method in detail.

\subsection{Motivation}

As aforementioned, it has been increasingly common to do LoRA fine-tuning in practice given the ever-growing size of machine learning models, as full fine-tuning might be too costly or infeasible in the first place~\cite{hu2021lora, hayou2024loraefficientlowrank, dettmers2024qlora, hyeon2021fedpara}.
However, we find that existing attack methods exhibit significantly diminished attack performance on the merged model when the malicious model is LoRA fine-tuned.
We showcase this observation with Table \ref{tab:fine-tuning_results}, where the state-of-the-art attack, BadMerging~\cite{zhang2024badmerging}, has a drop of 40-68\% in the attack success rate when switching from full fine-tuning to LoRA.

Our hypothesis on the cause of the degraded attack effect is that the relatively small weight updates introduced by LoRA may limit the fulfillment of the adversarial goal. 
This can be seen from Table \ref{tab:l2}, which displays the $\ell_2$ distance between the weights of the fine-tuned malicious model and the pre-trained model.

Attempting to enhance the attack performance under LoRA, our high-level idea is to amplify the weights that contribute to the malicious behavior.
To achieve this, we first notice a previous observation, which we refer to as the \textit{orthogonality finding}~\cite{liu2024loraasanattackpiercingllmsafety}. 
It says that after malicious fine-tuning, only certain layers of the model will primarily serve the attack purpose, while other layers are dedicated to maintaining the normal functionality of the model for downstream tasks (\ie, the malicious and benign layers within a model are almost orthogonal/disjoint with each other).

Inspired by this orthogonality finding, we propose LoBAM, a simple yet effective method that can achieve successful backdoor attack against the merged model with a LoRA-tuned malicious model.

\begin{table}[t]
\footnotesize
  \centering
  \caption{BadMerging attack success rate in on-task and off-task attack against model merging using full fine-tuning and LoRA fine-tuning.}
  \addtolength{\tabcolsep}{-1.9pt}
   
    {
    \begin{tabular}{ccccc}
    \toprule
     & full fine-tune & LoRA (r=4) & LoRA (r=8) & LoRA (r=16) \\ \midrule
     On-task & 98.56 & 46.78 & 57.33 & 58.62 \\
     Off-task & 98.27 & 30.42 & 35.30 & 41.86 \\ \bottomrule
    \end{tabular}%
    }
    \label{tab:fine-tuning_results}%
\end{table}

\begin{table}[t]
\footnotesize
  \centering
  \caption{$\ell_2$ distances between malicious models and pre-trained models under different fine-tuning methods.}
  \addtolength{\tabcolsep}{-1.9pt}
   
    {
    \begin{tabular}{ccccc}
    \toprule
     & full fine-tune & LoRA (r=4) & LoRA (r=8) & LoRA (r=16) \\ \midrule
     On-task & 129.37 & 4.22 & 5.30 & 7.99\\
     Off-task & 171.72 & 3.68 & 6.50 & 9.15 \\ \bottomrule
    \end{tabular}%
    }
    \label{tab:l2}%
\end{table}

\subsection{LoBAM}
\label{sec:4.2}

The key formulation of LoBAM is
\begin{equation}
\theta_{\text{upload}} = \lambda (\theta_{\text{malicious}} - \theta_{\text{benign}}) + \theta_{\text{benign}},
\label{eq:lobam}
\end{equation}
with the algorithmic pipeline shown in Algorithm \ref{alg:MaliciousModelConstruction}.

\myparatight{Obtaining $\theta_\text{malicious}$ and $\theta_\text{benign}$}
Here, the malicious model $\theta_\text{malicious}$ and the benign model $\theta_\text{benign}$ are both LoRA fine-tuned from the pre-trained model $\theta_\text{pre}$.
Specifically, $\theta_\text{malicious}$ is trained on poisoned images (clean images with triggers attached), with BadMerging ~\cite{zhang2024badmerging} being the malicious training objective.
Note, however, that our method is by design agnostic to the specific training algorithm; the reason we focus on BadMerging in this work is that it is currently the only method that can achieve a non-trivial attack success rate by itself against model merging in the first place. 
To train $\theta_\text{benign}$, just like any other benign users would do, we use standard cross-entropy loss to maximize the classification accuracy on the original clean dataset.

\myparatight{Constructing $\theta_\text{upload}$}
Unlike previous methods that naively upload the fine-tuned malicious model $\theta_\text{malicious}$ for model merging, LoBAM uniquely chooses to form the uploaded model using Equation \ref{eq:lobam}.
Intuitively, $\theta_{\text{malicious}}-\theta_{\text{benign}}$ isolates the key components that contribute to the attack goal based on the orthogonality finding~\cite{liu2024loraasanattackpiercingllmsafety}.
By scaling the difference with the factor $\lambda>1$, we are essentially amplifying the attack strength.
Finally, we treat the $\lambda$-scaled term as a residual and add it back to $\theta_\text{benign}$, anticipating that the weight distribution of the final model is close to that of the benign model, which can help maintain the normal downstream performance (without attacks). Figure \ref{fig:LoBAM} represents the illustration of LoBAM.

In the meantime, one may wonder if naively scaling the weights, \ie, $\theta_{\text{upload}} = \lambda\cdot \theta_{\text{malicious}}$, can boost the attack efficacy as well.
However, we posit that such a strategy will not selectively target the parameters linked to the malicious objective, and thus blindly amplifying all weights together would fail to enhance the malicious effects.
In fact, according to the empirical results shown in Table~\ref{tab:simple}, this naive scaling approach results in highly unsatisfactory attack success rates.
Therefore, we remark that our formulation in Equation \ref{eq:lobam} intelligently constructs the uploaded model.

\myparatight{Determining $\lambda$}
We propose a strategy to automatically and dynamically determine the value of $\lambda$, which is listed in Algorithm \ref{alg:BinarySearchLambda}.
In a nutshell, it iteratively adjusts $\lambda$ with binary search to ensure that the magnitude of $\theta_\text{upload}$ remains within a certain range. This regulation is crucial because if $\lambda$ is too small, the effectiveness of the attack diminishes. Conversely, if $\lambda$ is too large, it significantly deviates from the benign model, making it more likely to be detected. Our later experiments validated the necessity of this design.

\begin{algorithm}[t]
    \caption{LoBAM}
    \label{alg:MaliciousModelConstruction}
    \begin{algorithmic}[1]
        \renewcommand{\algorithmicrequire}{\textbf{Input:}}
        \renewcommand{\algorithmicensure}{\textbf{Output:}}
        \Require Pre-trained model $\theta_{\text{pre}}$, poisoned dataset $D_{\text{poisoned}}$, clean dataset $D_{\text{clean}}$ 
        \Ensure The model $\theta_{\text{upload}}$ that the attacker will upload for model merging
        
        \State \textbf{Step 1: Obtaining malicious fine-tuned model $\theta_{\text{malicious}}$ and benign fine-tuned model $\theta_{\text{benign}}$}
        \State Fine-tune $\theta_{\text{pre}}$ on $D_{\text{poisoned}}$ using LoRA to get $\theta_{\text{malicious}}$
        \State Fine-tune $\theta_{\text{pre}}$ on $D_{\text{clean}}$ using LoRA to get $\theta_{\text{benign}}$
        
        \State \textbf{Step 2: Construction of the uploaded model}

        \State Call Algorithm \ref{alg:BinarySearchLambda} to find the optimal $\lambda_{\text{val}}$
        
        \State $\theta_{\text{upload}} = \lambda_{\text{val}} \cdot (\theta_{\text{malicious}} - \theta_{\text{benign}}) + \theta_{\text{benign}}$
        
        \State \Return $\theta_{\text{upload}}$
    \end{algorithmic}
\end{algorithm}

\begin{algorithm}[t]
    \caption{Binary Search for $\lambda$ Adjustment}\label{alg:BinarySearchLambda}
    \begin{algorithmic}[1]
        \renewcommand{\algorithmicrequire}{\textbf{Input:}}
        \renewcommand{\algorithmicensure}{\textbf{Output:}}
        \Require Malicious model $\theta_{\text{malicious}}$, benign model $\theta_{\text{benign}}$, initial range $[\lambda_{\min}, \lambda_{\max}]$, initial value $\lambda_\text{val}=\frac{\lambda_{\min} + \lambda_{\max}}{2}$, tolerance $\epsilon$, initial PreDist = -1.
        \Ensure Optimal $\lambda_{\text{val}}$

        \While{$\lambda_{\max} - \lambda_{\min} > \epsilon$}
            \State $\theta_{\text{upload}} = \lambda_{\text{val}} \cdot (\theta_{\text{malicious}} - \theta_{\text{benign}}) + \theta_{\text{benign}}$
            
            \State $\text{Dist} = \|\theta_{\text{upload}} \|_2$
            
            \If{$\text{Dist} > \text{PreDist}$}
                \State $\lambda_{\max} = \lambda_{\text{val}}$
            \Else
                \State $\lambda_{\min} = \lambda_{\text{val}}$
            \EndIf
            
            \State $\lambda_{\text{val}} = \frac{\lambda_{\min} + \lambda_{\max}}{2}$
            \State $\text{PreDist} = \text{Dist}$
        \EndWhile
        
        \State \Return $\lambda_{\text{val}}$
    \end{algorithmic}
\end{algorithm}

%% file: 5_experiments.tex

\section{Experiments} \label{sec:exp}

\subsection{Experimental Setup}
\myparatight{Datasets}
In our experiments, we consider 10 widely used benchmarks, including CIFAR100~\cite{krizhevsky2009learning}, ImageNet100~\cite{deng2009imagenet}, SUN397~\cite{xiao2010sun}, GTSRB~\cite{stallkamp2011german}, SVHN~\cite{netzer2011reading}, MNIST~\cite{deng2012mnist}, Cars196~\cite{krause20133d}, EuroSAT~\cite{helber2019eurosat}, Pets~\cite{parkhi2012cats}, and STL10~\cite{coates2011analysis}. 

\myparatight{Compared attacks} 
We compare our method with  BadNets~\cite{gu2019badnetsidentifyingvulnerabilitiesmachine}, Dynamic Backdoor~\cite{salem2022dynamicbackdoorattacksmachine}, and BadMerging~\cite{zhang2024badmerging}.
Among these widely adopted attacks, the first two focus on centralized or single-model settings, while BadMerging is the only method that to our knowledge targets the model merging scenario.

\myparatight{Attack settings}
In our experiments, the attacker employed LoRA to fine-tune pre-trained models to execute both on-task and off-task attacks across all baselines as well as our LoBAM. We consider a model merging system in which each user fine-tunes a ViT-L/14 model~\cite{radford2021learningtransferablevisualmodels}. 
This model holds practical significance and real-world relevance for two primary reasons: First, its excellent performance has led to widespread adoption among users. Second, its substantial parameter count makes full fine-tuning computationally intensive, often forcing attackers to rely on LoRA as a resource-efficient alternative.

In each model merging case, we consider 5 benign users and 1 malicious user (the attacker), following BadMerging \cite{zhang2024badmerging}.
Each user has a different task/dataset at hand, and we consider 3 groups of random task assignments listed in Table~\ref{tab:dataset_combinations}.
In each combination, the first dataset represents the adversary task.
For an on-task attack, the adversary task itself is the target task, while in an off-task attack, the second dataset serves as the target attack. 
For instance, in combination `A', while CIFAR100 is the adversary task in both scenarios, the target task is CIFAR100 and SUN397 for on-task and off-task attack, respectively.
The targeted class within each task was randomly chosen from the corresponding dataset.
Notably, our setup closely follows previous works \cite{zhang2024badmerging} to ensure a straight and fair comparison.

\begin{table}[t]
\footnotesize
  \centering
  \caption{Task assignments in the model merging system considered in our experiments.}
  \addtolength{\tabcolsep}{-1.7pt}
    \resizebox{0.48\textwidth}{!}{  
    \begin{tabular}{cc}
    \toprule
    Combination & Datasets \\ \midrule
    A & CIFAR100, SUN397, EuroSAT, SVHN, Cars196, MNIST \\ 
    B & SVHN, Pets, EuroSAT, GTSRB, ImageNet100, STL10 \\ 
    C & MNIST, Cars196, ImageNet100, STL10, EuroSAT, GTSRB \\ \bottomrule
    \end{tabular}%
    }
    \label{tab:dataset_combinations}%
\end{table}

\begin{table*}[t]
\footnotesize
  \centering
  \caption{Attack success rate (\%) for different on-task attacks on dataset combination A, B, and C. SA, TA, Ties, and AM are four different algorithms for merging all users' models.}
  \addtolength{\tabcolsep}{-0.1pt}
  \resizebox{0.99\linewidth}{!}{%
  \begin{tabular}{l c c c c c c c c c c c c}
    \toprule
    \multirow{2}{*}{Attack} & \multicolumn{4}{c}{A} & \multicolumn{4}{c}{B} & \multicolumn{4}{c}{C} \\ \cmidrule(lr){2-5} \cmidrule(lr){6-9} \cmidrule(lr){10-13}
    & SA & TA & Ties & AM & SA & TA & Ties & AM & SA & TA & Ties & AM \\ \midrule
    BadNets ~\cite{gu2019badnetsidentifyingvulnerabilitiesmachine}           & 0.35   & 1.86   & 0.83  & 1.29        & 0.84 & 0.45 & 0.18 & 0.71            & 0.21  & 0.09 & 0.53 & 0.87       \\
    Dynamic Backdoors ~\cite{salem2022dynamicbackdoorattacksmachine} & 1.26   & 2.57   & 1.34  & 2.82        & 3.73 & 1.98 & 1.32 & 2.37            & 2.71  & 1.84 & 3.45 & 1.83       \\
    BadMerging ~\cite{zhang2024badmerging}      & 53.78  & 57.33  & 46.97 & 36.02       & 57.63 & 40.36 & 39.84 & 19.50         & 51.82 & 53.65 & 65.70 & 10.20 \\
    \rowcolor{gray!15}LoBAM             & \textbf{98.69}  & \textbf{99.40}  & \textbf{98.12} & \textbf{74.51}       & \textbf{99.31} & \textbf{98.77} & \textbf{99.94} & \textbf{85.86}         & \textbf{96.38} & \textbf{99.21} & \textbf{98.42} & \textbf{73.66} \\
    \bottomrule
  \end{tabular}
  }
  \label{tab:lora_comparison_on}
\end{table*}

\myparatight{Model Merging Algorithms}
In our experiments, we consider the following model merging algorithms.

\textit{Simple Averaging (SA)~\cite{wortsman2022model}:}
SA computes the merged weights as the element-wise arithmetic mean of
the weights of all other models. Suppose there are $N$ models and the $i$-th model is $\theta_i$, then the weight updates between the merged model and the pre-trained model $\Delta \theta_{\text{merged}}$ is calculated as $
\Delta\theta_{\text{merged}} = \frac{1}{N} \sum_{\substack{i=1}}^{N} \Delta \theta_i$.

\textit{Task Arithmetic (TA):~\cite{ilharco2022editing}:}
TA is similar to the SA in that it makes every task vector have the same contribution to the merged model. The only difference is that TA further uses a scaling factor $k$, where $\Delta \theta_{\text{merged}} = k\cdot\sum_{\substack{i=1}}^{N} \Delta \theta_i$.

\textit{Ties Merging (Ties)~\cite{yadav2024ties}:}
Different from TA, Ties Merging takes the disjoint mean of each weight update, $\Phi(\Delta \theta_i)$, and scales and combines them.
Essentially, $\Delta \theta_{\text{merged}} = \alpha \cdot\sum_{\substack{i=1}}^{N} \Phi(\Delta \theta_i)$, where $\alpha$ is a scaling term.

\textit{AdaMerging (AM)~\cite{yang2023adamerging}:} In AdaMerging, it learns a unique scaling factor $k_i$ for each model update $\Delta \theta_{i}$, \ie, $\Delta \theta_{\text{merged}} = \sum_{\substack{i=1}}^{N} k_i \cdot \Delta \theta_{i}$. Specifically, the scaling factors $k_i$ are learned through an unsupervised entropy minimization objective.
Since it involves a learning process, AdaMerging is significantly more time-consuming compared to other merging algorithms.

\myparatight{Parameter setting} 
When the attacker constructs the malicious model, we set $r=8$ for LoRA and $\lambda_\text{min}=4$ and $\lambda_\text{max}=10$ for Algorithm \ref{alg:BinarySearchLambda}.
Later we will show the results under various $r$ and $\lambda$ for our method.

\myparatight{Metric}
We use attack success rate (ASR) as the metric to measure the effectiveness of the attack. Specifically, ASR measures the proportion of trigger-attached malicious inputs that are classified by the compromised model as the target class as the attacker intended. A high ASR indicates a highly effective attack.

\begin{table*}[t]
\footnotesize
  \centering
  \caption{Attack success rate (\%) for different off-task attacks on dataset combination A, B, and C. SA, TA, Ties, and AM are four different algorithms for merging all users' models.}
  \addtolength{\tabcolsep}{-0.1pt}
  \resizebox{0.99\linewidth}{!}{%
  \begin{tabular}{l c c c c c c c c c c c c}
    \toprule
    \multirow{2}{*}{Attack} & \multicolumn{4}{c}{A} & \multicolumn{4}{c}{B} & \multicolumn{4}{c}{C} \\ \cmidrule(lr){2-5} \cmidrule(lr){6-9} \cmidrule(lr){10-13}
    & SA & TA & Ties & AM & SA & TA & Ties & AM & SA & TA & Ties & AM \\ \midrule
    BadNets ~\cite{gu2019badnetsidentifyingvulnerabilitiesmachine}         & 0.05   & 0.17   & 0.13  & 0.12        & 0.35 & 0.22 & 0.51 & 0.08            & 0.13  & 0.01 & 0.07 & 0.24       \\
    Dynamic Backdoors ~\cite{salem2022dynamicbackdoorattacksmachine} & 0.32   & 1.28   & 0.45  & 0.25        & 1.06 & 0.74 & 2.23 & 1.45            & 1.18  & 0.47 & 2.34 & 1.96       \\
    BadMerging ~\cite{zhang2024badmerging}      & 34.84  & 35.30  & 45.14 & 35.61       & 47.24 & 32.33 & 55.62 & 19.59         & 44.02 & 50.01 & 48.53 & 16.31 \\
    \rowcolor{gray!15}LoBAM             & \textbf{97.47}  & \textbf{98.97}  & \textbf{99.65} & \textbf{71.43}       & \textbf{99.81} & \textbf{99.94} & \textbf{100} & \textbf{89.92}         & \textbf{95.93} & \textbf{97.23} & \textbf{94.88} & \textbf{75.25} \\
    \bottomrule
  \end{tabular}
  }
  \label{tab:lora_comparison_off}
\end{table*}

\subsection{Experimental Results}
\myparatight{LoBAM consistently outperforms all baselines in on-task and off-task attack scenarios} 
We evaluate the effectiveness of on-task and off-task attacks of LoBAM, alongside several baseline methods, across four commonly used model merging algorithms. 
The on-task attack results are presented in Table~\ref{tab:lora_comparison_on}, and the off-task attack results are shown in Table~\ref{tab:lora_comparison_off}. 
The task assignment or dataset combination is represented by ``A, B, C.''

From the results, we observe that backdoor attacks originally designed for single models, such as BadNets and Dynamic Backdoors, have minimal effect in this setting, achieving attack success rates (ASR) below 10\%.
While BadMerging (specifically tailored for model merging) demonstrates excellent performance under full fine-tuning (recall Table~\ref{tab:fine-tuning_results}), its effectiveness diminishes significantly under LoRA fine-tuning context, where the attack success rate typically ranges from 30\% to 50\%.
In contrast, our LoBAM achieves around 98\% attack success rate in most cases, highlighting its superior efficacy.

\begin{figure}[!t]
	\centering
	\includegraphics[width=0.95\linewidth]{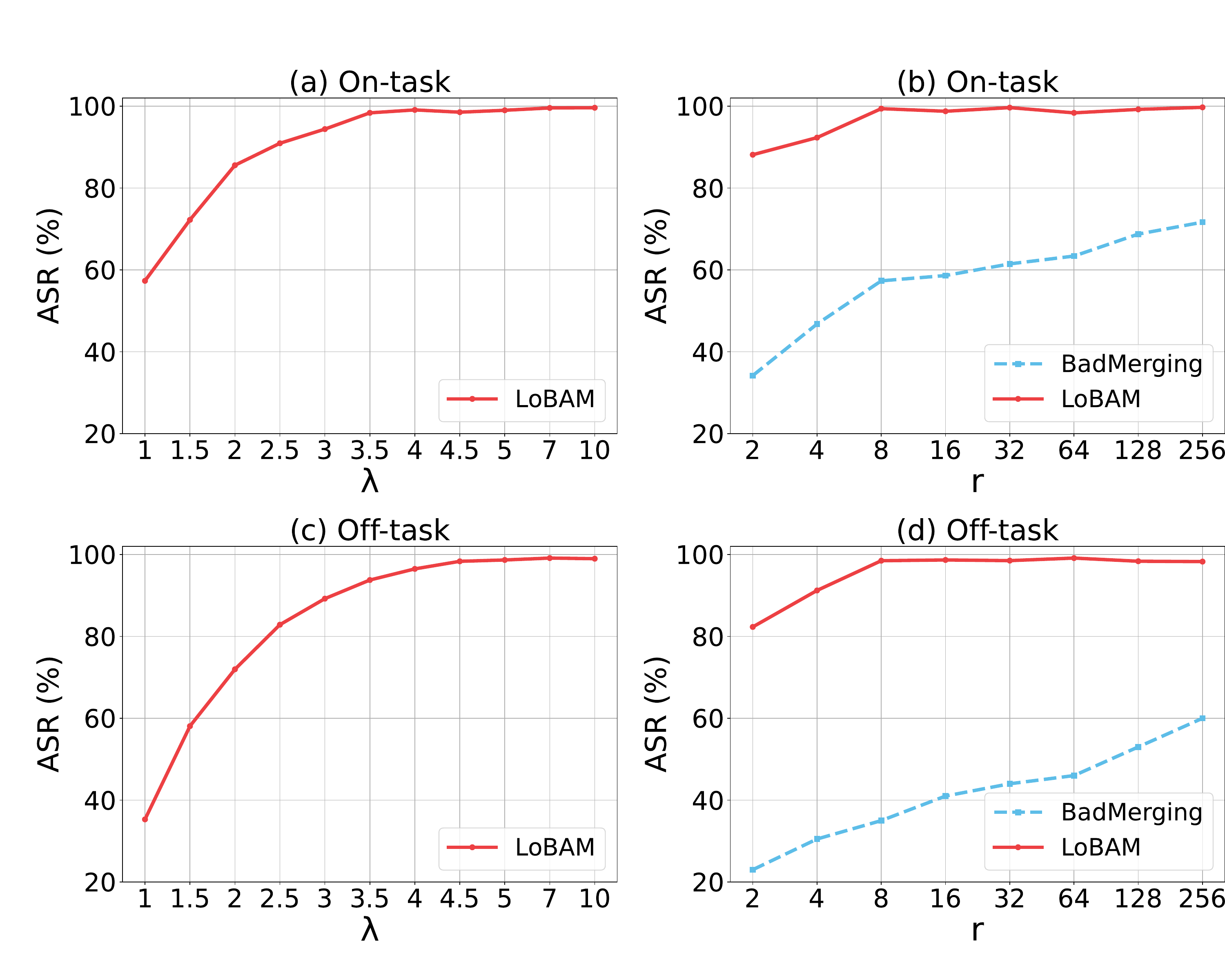}
	\caption{Result of ablation studies on $r$ and $\lambda$, where TA is the merging algorithm.}
		\label{fig:ablation}
\end{figure}

\myparatight{Study on the impact of $r$}
In the LoRA fine-tuning process, the parameter $r$ signifies the number of trainable parameters. This section explores the effects of varying $r$ values by setting it to 2, 4, 8, 16, 32, 64, 128, and 256. 
We assess the attack success rate with the dataset combination A and use Task Arithmetic as the merging algorithm. Figure \ref{fig:ablation} (b) and (d) demonstrate the experimental results in the on-task and the off-task scenario, respectively. 
It is evident that the attack success rate for both BadMerging and LoBAM generally increases with larger $r$ values. 
This trend is due to the insufficient number of parameters updated during fine-tuning when $r$ is small (i.e., $r = 2$ or $r = 4$). 
Nevertheless, even when $r$ is small, LoBAM still achieves commendable attack performance with an attack success rate exceeding 80\%. 
When $r$ is increased to 8, LoBAM’s ASR already surpasses 98\%, avoiding the need to further increase $r$ which incurs extra computational cost. 

\myparatight{Study on the impact of $\lambda$}
In our method, the parameter $\lambda$ represents the amplification factor used to enhance the influence of the malicious model in the model merging system. Intuitively, a larger $\lambda$ is considered advantageous, and as $\lambda$ surpasses a certain threshold, the effectiveness of the attack may be saturated. To examine the precise impact of different $\lambda$ values, we vary $\lambda$ from 1 to 10 and measure the attack success rate. 
We conduct experiments again with combination A and use Task Arithmetic as the merging algorithm. The results of on-task and off-task scenarios are presented in Figure \ref{fig:ablation} (a) and (c), respectively.

It is evident that when $\lambda$ is 1, which essentially degenerates to not applying our LoBAM method, the attack success rate is notably low. 
As $\lambda$ increases, the attack success rate rises, ultimately reaching saturation at approximately $\lambda = 3.5$ for the on-task scenario and $\lambda = 4.5$ for the off-task scenario.
Further increasing $\lambda$ to a large value, say 8, 10, or 15, will make the resulting model significantly different from the pre-trained model, in terms of the $\ell_2$ distance between the model weights shown in Table \ref{tab:benign_lobam}.
More specifically, when $\lambda$ is large, the $\ell_2$ distance between the uploaded malicious model and the pre-trained model is much larger than that between the benign model and the pre-trained model, meaning that a simple distance thresholding might detect the malicious one and exclude it from model merging, preventing a successful attack.
However, by dynamically determining a $\lambda$ within a specific range as our method does, we can ensure 1) a decent attack success rate, and 2) that the modifications to the pre-trained model are similar to those seen in benign models (at least in terms of distance).

\begin{table}[h!]
    \centering
     \caption{$\ell_2$ distances between LoBAM malicious models and pre-trained models under different $\lambda$ along with the $\ell_2$ distances between benign users' models and pre-trained models.}
    \resizebox{0.9\linewidth}{!}{%
    \begin{tabular}{c c c c c c }
        \toprule
        \multirow{2}{*}{Benign} & \multicolumn{5}{c}{LoBAM} \\ \cmidrule{2-6}
        & $\lambda=4$ & $\lambda=6$ & $\lambda=8$ & $\lambda=10$ & $\lambda=15$ \\ \midrule
        61.82 & 34.59 & 57.03 & 79.54 & 102.08 & 158.45 \\ \bottomrule
    \end{tabular}
    }
    \label{tab:benign_lobam}
\end{table}

\begin{table}[h!]
\centering
\caption{Attack success rate (\%) on on-task and off-task attack scenarios with different N using the TA merging algorithm.}
\resizebox{0.99\linewidth}{!}{%
\begin{tabular}{c cc cc cc}
\toprule
\multirow{2}{*}{N} & \multicolumn{2}{c}{A} & \multicolumn{2}{c}{B} & \multicolumn{2}{c}{C} \\ \cmidrule(lr){2-3} \cmidrule(lr){4-5}  \cmidrule(lr){6-7}
                   & on-task & off-task      & on-task & off-task      & on-task & off-task      \\ \midrule
2                  & 100  & 99.61         & 99.05   & 99.51         & 99.38   & 99.47         \\
4                  & 99.57   & 98.32         & 98.81   & 99.13         & 99.74   & 97.82         \\
6                  & 99.40   & 98.97         & 98.70   & 99.94         & 99.21   & 97.23         \\
8                  & 95.38   & 98.94         & 96.41   & 92.37         & 93.85   & 94.53         \\ \bottomrule
\end{tabular}%
}
\label{tab:model_performance}
\end{table}

\myparatight{Study on the impact of $N$}
$N$ denotes the total number of models to be merged within the model merging system. By default, $N$ is set to 6. Here, we explore the impact of varying $N$ by setting it to 2, 4, 6, and 8. For $N = 2$ and $N = 4$, we select the first 2 and first 4 datasets from each combination, respectively. Referring to Table \ref{tab:model_performance}, which presents the performance of LoBAM on the Task Arithmetic, we observe that LoBAM consistently achieves great results across different values of $N$.

\myparatight{Study on naively scaling the malicious weights}
As mentioned earlier in Section \ref{sec:4.2}, the most straightforward attempt to amplify the malicious impact is naively scaling the malicious weights by setting $\theta_{\text{upload}} = \lambda\cdot \theta_{\text{malicious}}$. This section evaluates the effectiveness of such an approach by adjusting $\lambda$ to 1, 1.5, 2, 3, 4, 5, and 6, and then testing the corresponding attack success rate. The results in Table~\ref{tab:simple} indicate that this strategy is extremely ineffective, with ASR falling below 1\% when $\lambda$ exceeds 2. This ineffectiveness arises because the scaling approach does not selectively target parameters associated with the malicious objectives; hence, indiscriminately amplifying all weights simultaneously fails to achieve excellent attack effect.

\begin{table}[t]
\footnotesize
  \centering
  \caption{Attack success rate (\%) on on-task and off-task scenarios under different $\lambda$ when naively using $\theta_{\text{upload}} = \lambda\cdot \theta_{\text{malicious}}$, where TA is the merging algorithm.}
  \addtolength{\tabcolsep}{0.5pt}
   
    {
    \begin{tabular}{c c c c c c c c}
    \toprule
     $\lambda$ & 1 & 1.5 & 2 & 3 & 4 & 5 & 6 \\ \midrule
     On-task & 57.33 & 7.69 & 0.07 & 0.04 & 0.02 & 0.01 & 0\\
     Off-task & 35.30 & 0.28 & 0.02 & 0.06 & 0.01 & 0.01 & 0.02 \\ \bottomrule
    \end{tabular}%
    }
    \label{tab:simple}
\end{table}

\myparatight{Study on the impact of different targeted class}
We also evaluated the effectiveness of the LoBAM attack across various target classes within the `A' Combination. For both on-task and off-task attacks, we selected three distinct target classes and measured the attack success rate of the LoBAM attack. The results, presented in Table \ref{tab:different class}, demonstrate that LoBAM consistently achieves high performance across diverse target classes.

\begin{table}[t!]
    \centering
     \caption{Attack success rate (\%) on on-task and off-task scenarios for different target classes.}
    \resizebox{0.99\linewidth}{!}{%
    \begin{tabular}{l>{\hspace*{2mm}}c>{\hspace*{2mm}}c>{\hspace*{2mm}}c>{\hspace*{2mm}}c>{\hspace*{2mm}}c>{\hspace*{2mm}}c}
        \toprule
        & \multicolumn{3}{c}{On-task} & \multicolumn{3}{c}{Off-task} \\
        \cmidrule(lr){2-4} \cmidrule(lr){5-7}
        & {\small\texttt{Mountain}} & {\small\texttt{Bed}} & {\small\texttt{Rose}} & {\small\texttt{Arch}} & {\small\texttt{Canyon}} & {\small\texttt{Waterfall}} \\
        \midrule
        SA & 99.74 & 98.56 & 98.21 & 97.12 & 98.20 & 96.27 \\
        TA & 97.01 & 98.35 & 98.56 & 99.24 & 95.86 & 97.37 \\
        Ties & 99.22 & 98.78 & 99.31 & 97.73 & 98.16 & 98.55 \\
        AM & 71.67 & 79.42 & 73.83 & 81.05 & 74.25 & 76.10 \\
        \bottomrule
    \end{tabular}%
    }
    \label{tab:different class}
\end{table}

\myparatight{Study on the impact of benign users using LoRA}
In our default setting, we assume that only the attacker, constrained by limited computational resources or aiming for greater efficiency, opts to use LoRA for model fine-tuning, while all benign users employ full fine-tuning. However, in reality, it is possible that benign users might also choose to fine-tune their models using LoRA. Therefore, in this section, we examine the efficacy of LoBAM when all benign users utilize LoRA for fine-tuning. The experimental results, shown in Table~\ref{tab:benign_lora}, indicate that LoBAM maintains excellent performance when benign users adopt LoRA for model fine-tuning.

\myparatight{Study on the impact of different models}
In the default setting, we primarily focus on scenarios where all users use the ViT-L/14 model due to the reasons mentioned above. However, there are cases where users employ smaller models, such as ViT-B/32 or ViT-B/16, in the model merging system. To evaluate the effectiveness of LoBAM in such scenarios, we conducted experiments accordingly. As demonstrated in Table~\ref{tab:32} and Table~\ref{tab:16}, the results reveal that LoBAM maintains excellent performance across varying models.

\begin{table}[t]
\centering
\caption{Attack success rate (\%) on on-task and off-task scenarios when all benign users use LoRA fine-tuning.}
\resizebox{0.99\linewidth}{!}{%
\begin{tabular}{c cc cc cc}
\toprule
\multirow{2}{*}{} & \multicolumn{2}{c}{A} & \multicolumn{2}{c}{B} & \multicolumn{2}{c}{C} \\ \cmidrule(lr){2-3} \cmidrule(lr){4-5}  \cmidrule(lr){6-7} 
                   & on-task & off-task      & on-task & off-task      & on-task & off-task      \\ \midrule
SA                  & 99.43  & 99.86       & 100   & 99.73         & 99.09   & 99.67         \\
TA                 & 99.98   & 100         & 99.75   & 99.08        & 100   & 99.51         \\
Ties                  & 100   & 99.63         & 100   & 99.76         & 100   & 99.84         \\
AM&   100 &    99.45    &  99.94  &  100     & 98.55  &     98.31     \\ \bottomrule
\end{tabular}%
}
\label{tab:benign_lora}
\end{table}

\begin{table}[t]
\centering
\caption{Attack success rate (\%) on on-task and off-task scenarios on dataset combination A, B, and C when all users use ViT-B/32 in the model merging system. SA, TA, Ties, and AM are four different algorithms.}

\resizebox{0.99\linewidth}{!}{%
\begin{tabular}{c cc cc cc}
\toprule
\multirow{2}{*}{} & \multicolumn{2}{c}{A} & \multicolumn{2}{c}{B} & \multicolumn{2}{c}{C} \\ \cmidrule(lr){2-3} \cmidrule(lr){4-5}  \cmidrule(lr){6-7} 
                   & on-task & off-task      & on-task & off-task      & on-task & off-task      \\ \midrule
SA                  & 98.26  & 99.71      & 98.45   & 99.85         & 100   & 97.32         \\
TA                 & 100   & 99.93         & 100   & 100        & 99.31   & 99.82         \\
Ties                  & 98.05   & 99.57       & 97.52   & 99.94         & 98.16   & 100         \\
AM&   99.84 &    98.47    &  100  &  98.77     & 99.55  &    99.42     \\ \bottomrule
\end{tabular}%
}
\label{tab:32}
\end{table}

\subsection{Safety Detection and Defense}
\myparatight{Safety Detection}

To ensure that the malicious model generated by LoBAM remains undetectable when uploaded to an open platform and can be selected for model merging, we perform t-SNE analysis \cite{van2008visualizing, chan2018t} on both benign and malicious models. t-SNE excels at preserving the original data distributions in a lower-dimensional space, making it a superior choice for identifying and defending against malicious activity according to previous research \cite{zhang2022pipattack, 10.1145/3638530.3654258, manikandan2024multiagent}.

We analyze a total of 80 models—60 benign models fine-tuned on various tasks and 20 malicious models created using LoBAM, also derived from diverse datasets. We apply t-SNE to reduce the dimensionality of model parameters from each layer to three dimensions for visualization purposes. The visualization results for all the layers are depicted in Figure~\ref{tsne} in the Appendix, demonstrating that the parameters of benign and malicious models are indistinguishable within the latent space. This finding provides compelling evidence of the robust concealment capabilities of our proposed attack, confirming its evasion from safety detection.

\myparatight{Defense}
After the model merging process, the model merging creator can employ various defense methods on the merged model to mitigate the attack~\cite{zhang2024badmerging}. To assess whether our attack remains effective in this context, we conduct experiments using well-known defense methods such as Neural Cleanse \cite{wang2019neural} and Scale-up \cite{guo2023scale}. The results presented in Table~\ref{tab:nc} and Table~\ref{tab:scale} demonstrate that even when defense methods are employed, our attack continues to exhibit strong effectiveness.

\begin{table}[t]
\centering
\caption{Attack success rate (\%) on on-task and off-task scenarios on dataset combination A, B, and C when all users use ViT-B/16 in the model merging system. SA, TA, Ties, and AM are four different algorithms.}
\resizebox{0.99\linewidth}{!}{%
\begin{tabular}{c cc cc cc}
\toprule
\multirow{2}{*}{} & \multicolumn{2}{c}{A} & \multicolumn{2}{c}{B} & \multicolumn{2}{c}{C} \\ \cmidrule(lr){2-3} \cmidrule(lr){4-5}  \cmidrule(lr){6-7} 
                   & on-task & off-task      & on-task & off-task      & on-task & off-task      \\ \midrule
SA                  & 100  & 99.01       &  99.82  & 99.24         & 99.30   & 99.61         \\
TA                 & 99.35   & 98.11         & 100   & 99.50        & 98.83   & 100         \\
Ties                  & 99.59   & 99.34        & 99.04   & 100         & 99.85   & 99.13         \\
AM&   99.22 &    99.74    &  99.52  &  99.75     & 97.59  &     99.18     \\ \bottomrule
\end{tabular}%
}
\label{tab:16}
\end{table}

%% file: 6_conclusion.tex

\section{Conclusion} \label{sec:conclusion}

In this paper, we discovered that existing backdoor attacks on model merging become ineffective due to attackers' limited computational resources and the resulting reliance on LoRA for fine-tuning pre-trained models. Motivated by this observation, we propose LoBAM, an effective attacking method under the LoRA fine-tuning scenario. LoBAM strategically combines the weights of a malicious and a benign model—each LoRA fine-tuned by the attacker—to amplify attack-relevant components, enhancing the model's malicious efficacy when deployed in model merging. Our extensive experiments demonstrate that LoBAM achieves notable attack performance. Additionally, our method exhibits excellent stealthiness, making it difficult to detect and defend against using conventional methods. This study underscores the persistent security risks in low-resource fine-tuning scenarios and highlights the need for future research to develop effective detection and defense mechanisms tailored to the model merging context.

\begin{table}[t]
\footnotesize
\centering
\caption{The result of the  Neural Cleanse defense on on-task and off-task scenarios
on dataset combination A, B, and C. The attack is successful against the defense if the score is less than 2.}
\begin{tabular*}{0.45\textwidth}{@{\extracolsep{\fill}} ccc}
\toprule
 & On-task & Off-task \\  
\midrule
A & 1.33 & 1.09 \\  
B & 1.57 & 0.83 \\  
C & 1.04 & 1.46 \\  
\bottomrule
\end{tabular*}
\label{tab:nc}
\end{table}

\begin{table}[t]
\footnotesize
\centering
\caption{The result of the Scale-up defense. The attack is considered successful against the defense if the score is less than 0.9.}
\begin{tabular*}{0.45\textwidth}{@{\extracolsep{\fill}} ccc}
\toprule
 & On-task & Off-task \\  
\midrule
A & 0.56 & 0.39 \\  
B & 0.32 & 0.64 \\  
C & 0.48 & 0.37 \\  
\bottomrule
\end{tabular*}
\label{tab:scale}
\end{table}

%% file: appendix.tex
\appendix

\clearpage
\begin{figure*}[!t]
	\centering
	\subfloat[Layer1]{\includegraphics[width=0.2\textwidth]{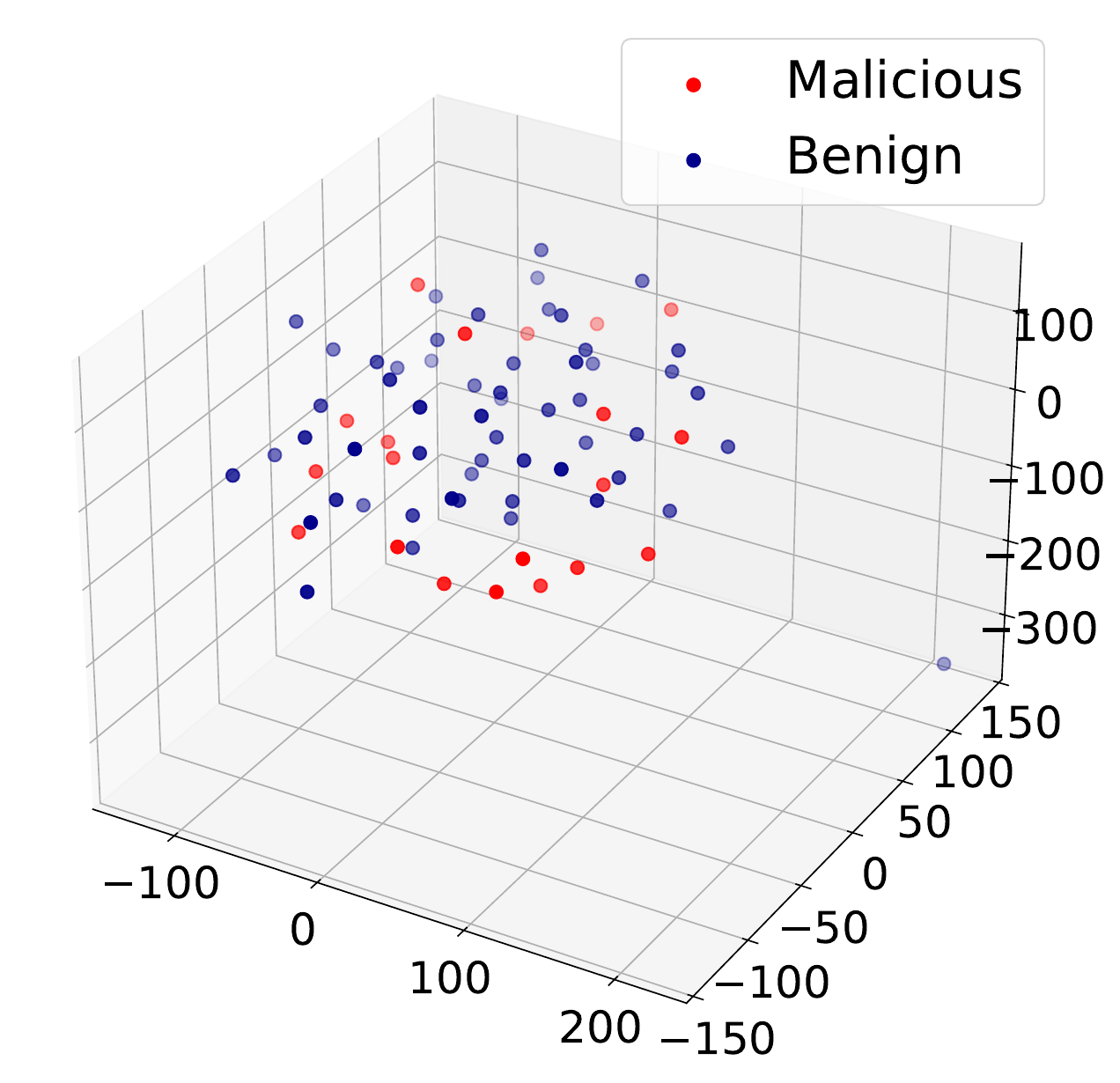}}
	\subfloat[Layer2]{\includegraphics[width=0.2\textwidth]{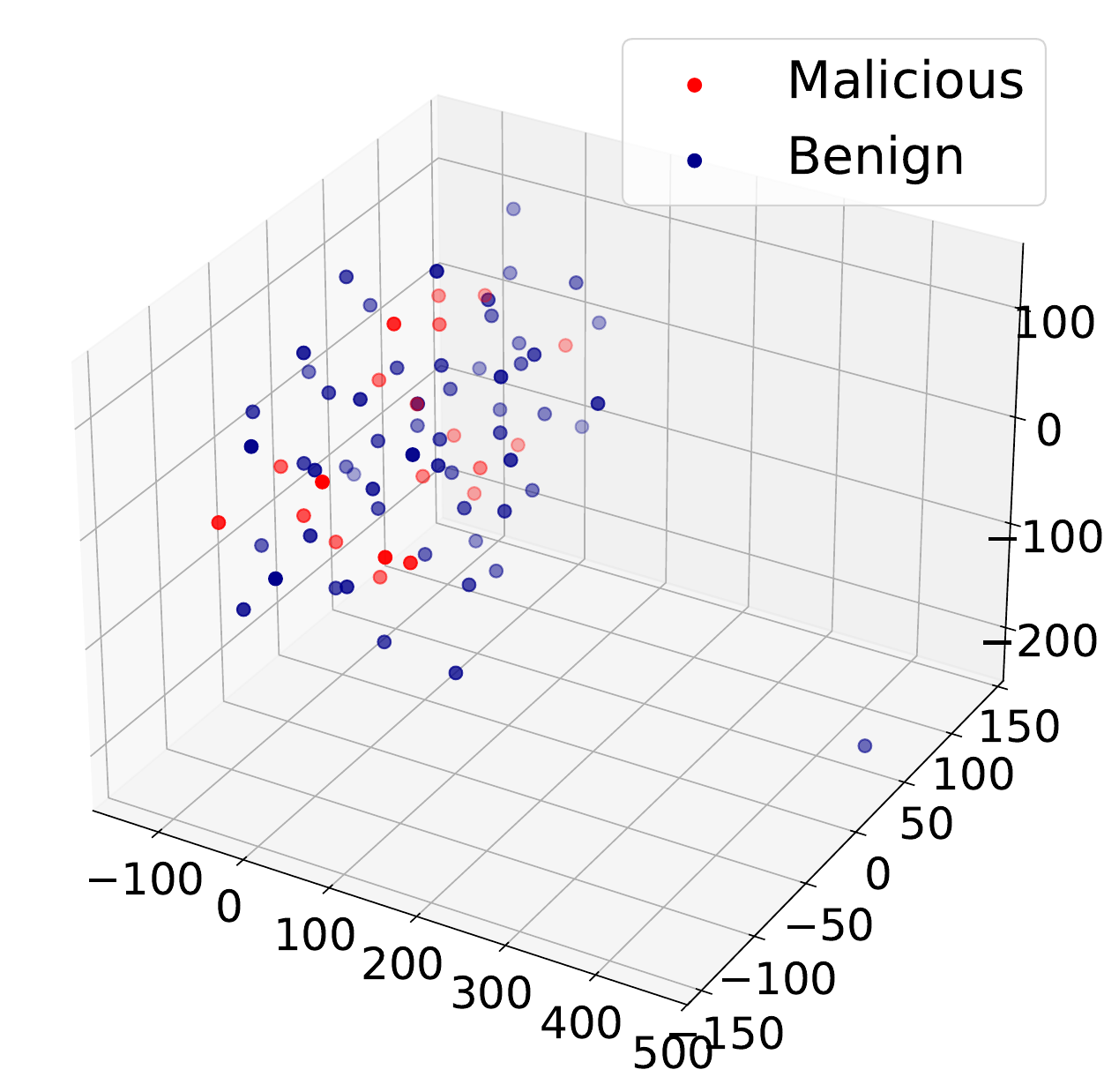}}
	\subfloat[Layer3]{\includegraphics[width=0.2\textwidth]{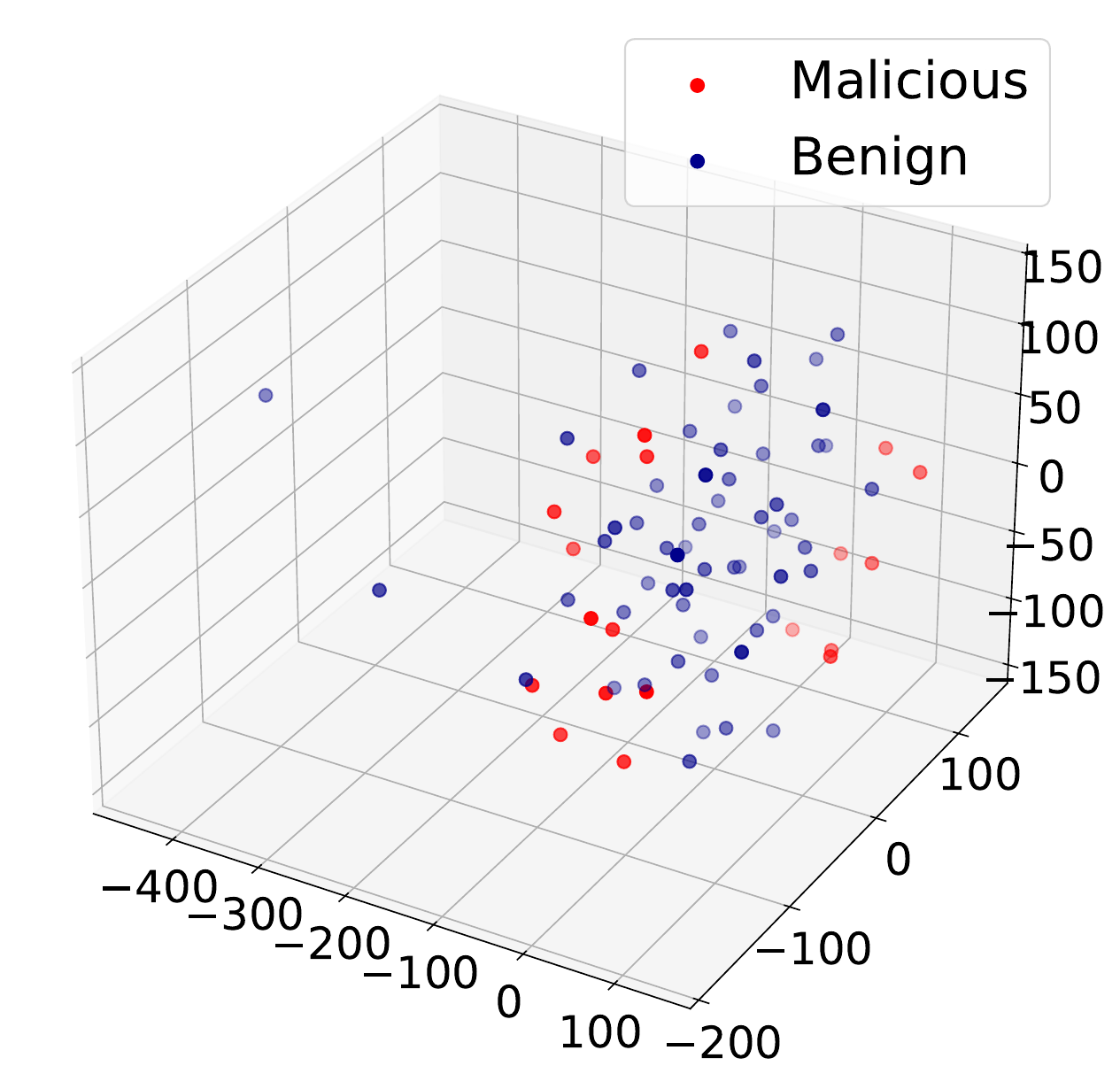}}
	\subfloat[Layer4]{\includegraphics[width=0.2\textwidth]{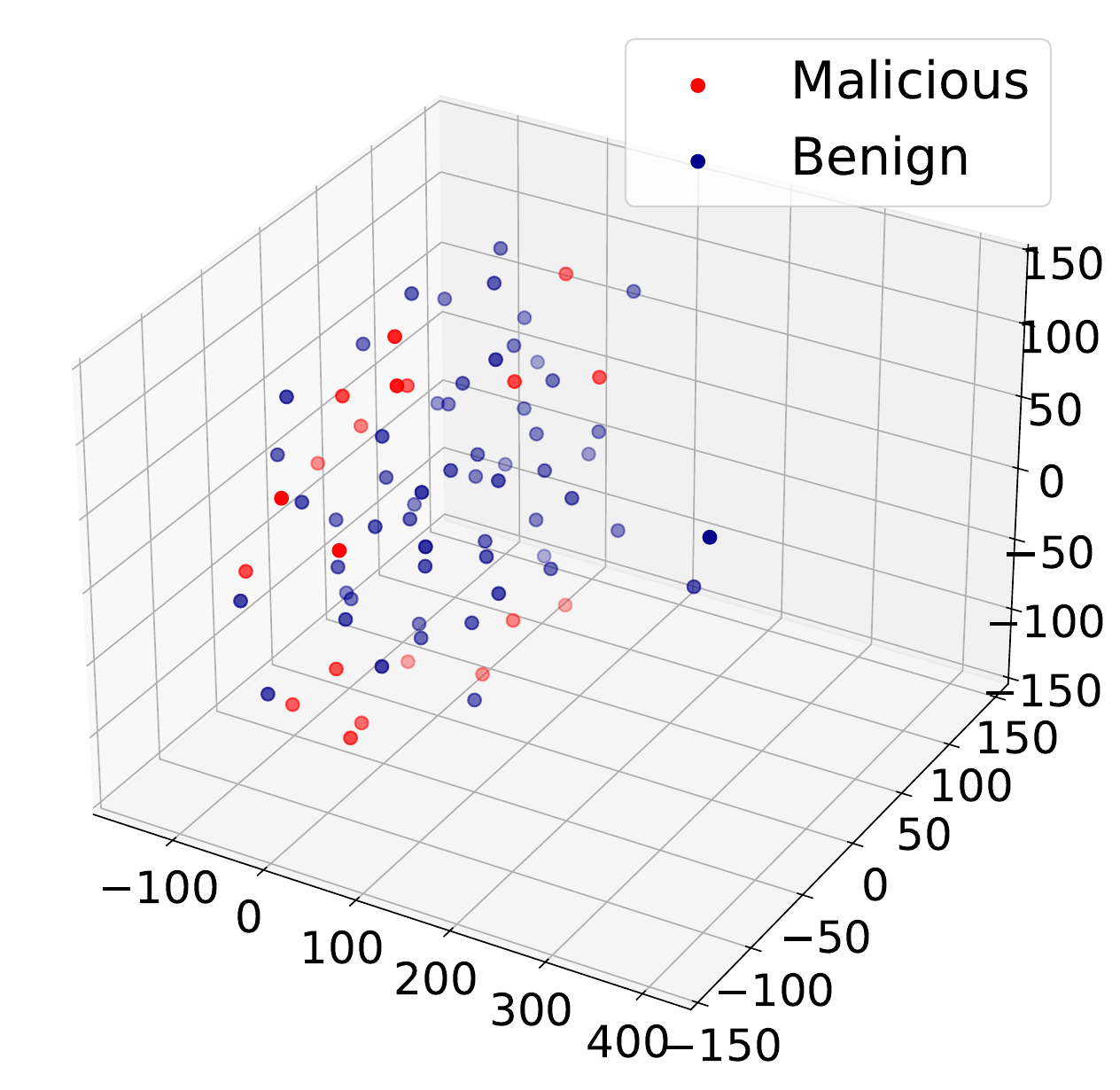}}
	\subfloat[Layer5]{\includegraphics[width=0.2\textwidth]{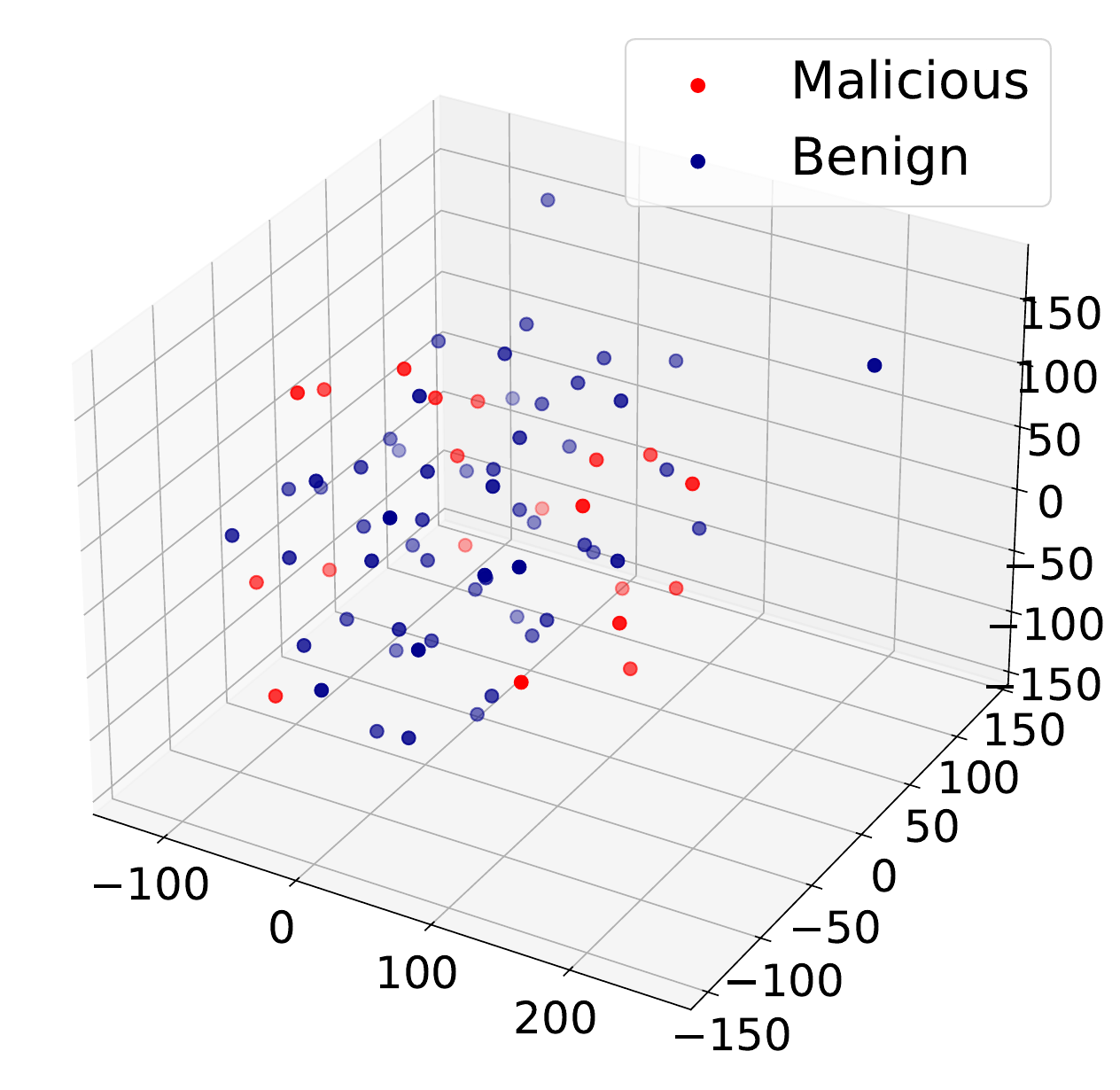}}\\
	\subfloat[Layer6]{\includegraphics[width=0.2\textwidth]{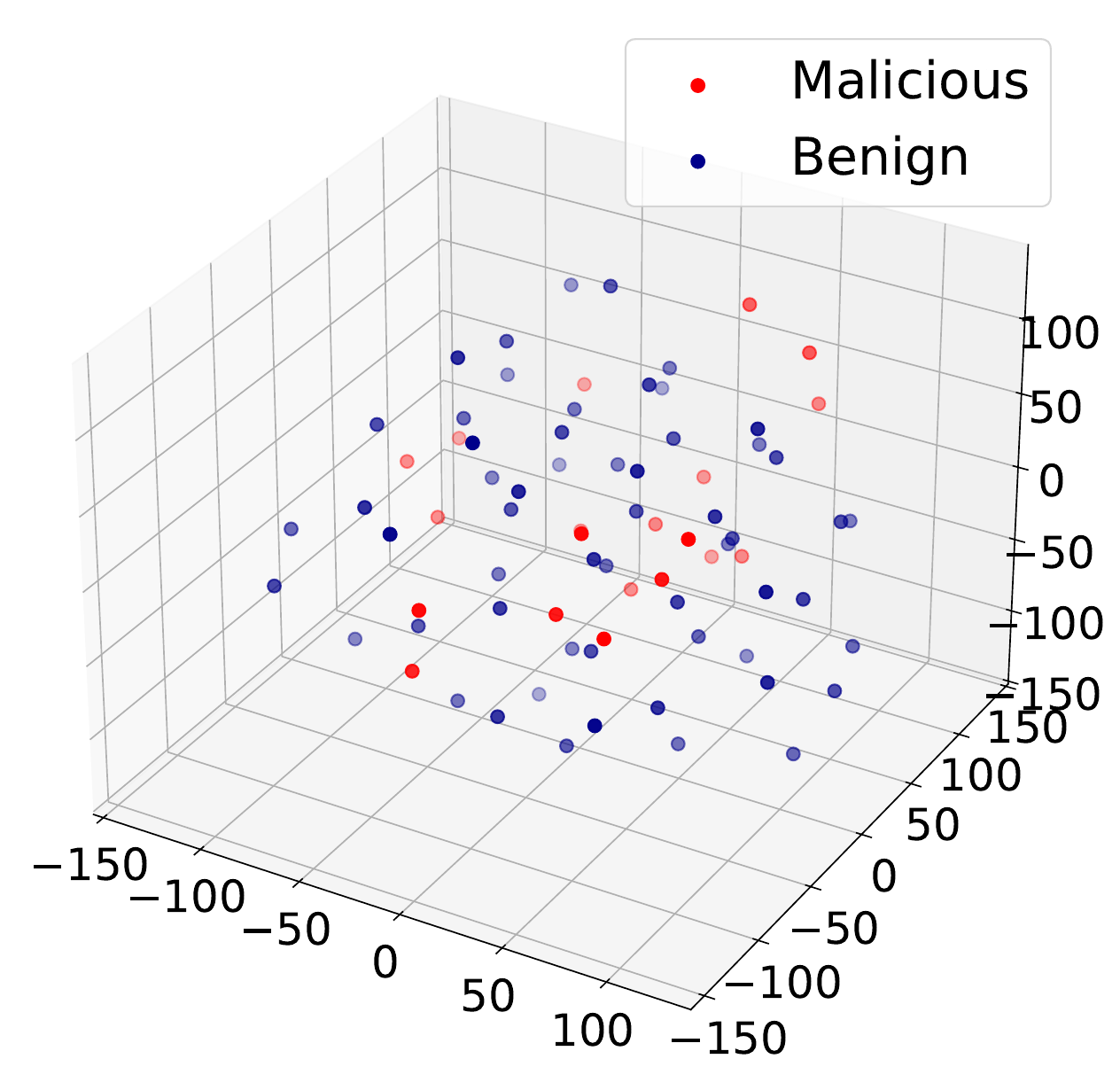}}
	\subfloat[Layer7]{\includegraphics[width=0.2\textwidth]{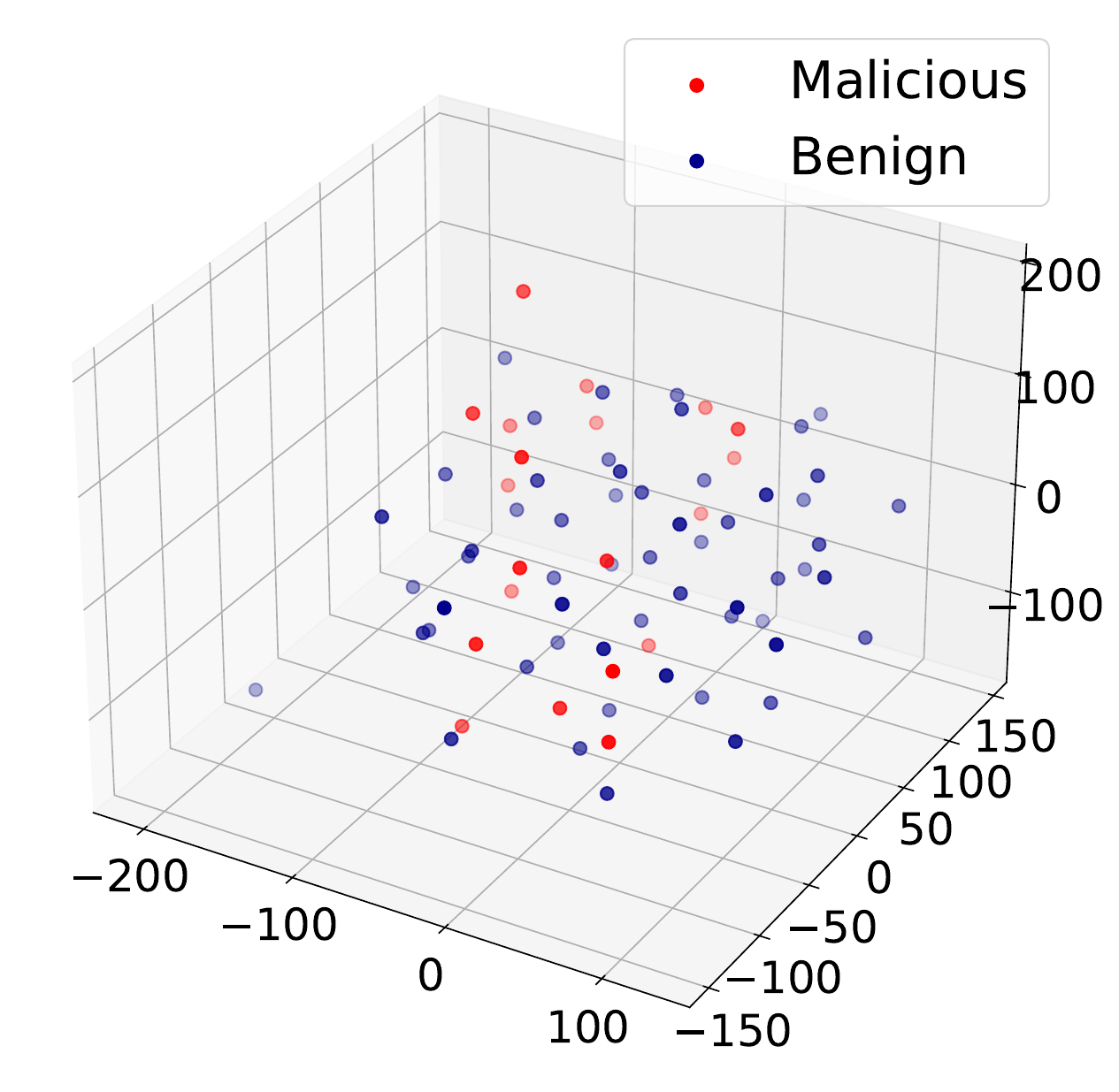}}
	\subfloat[Layer8]{\includegraphics[width=0.2\textwidth]{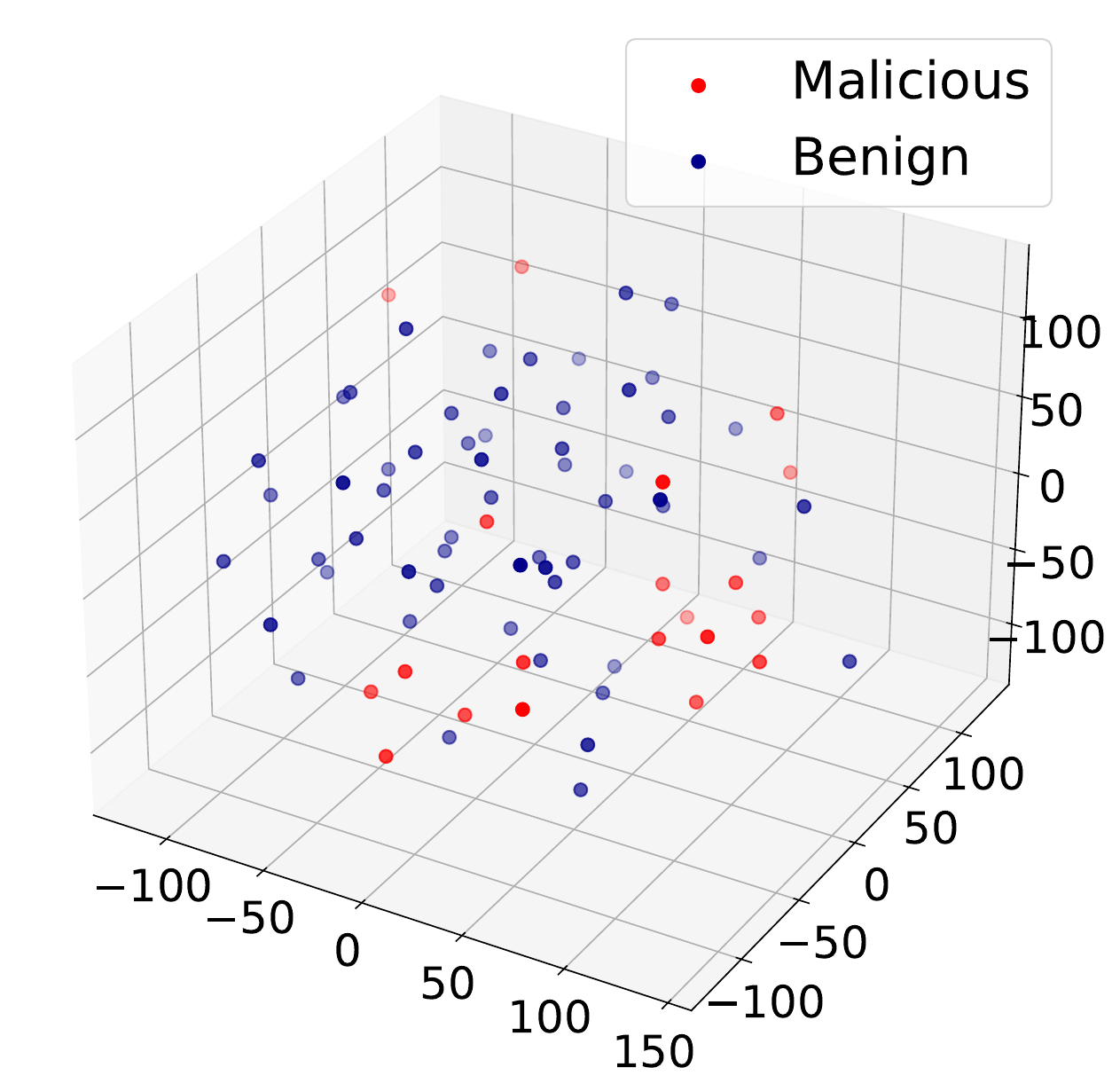}}
	\subfloat[Layer9]{\includegraphics[width=0.2\textwidth]{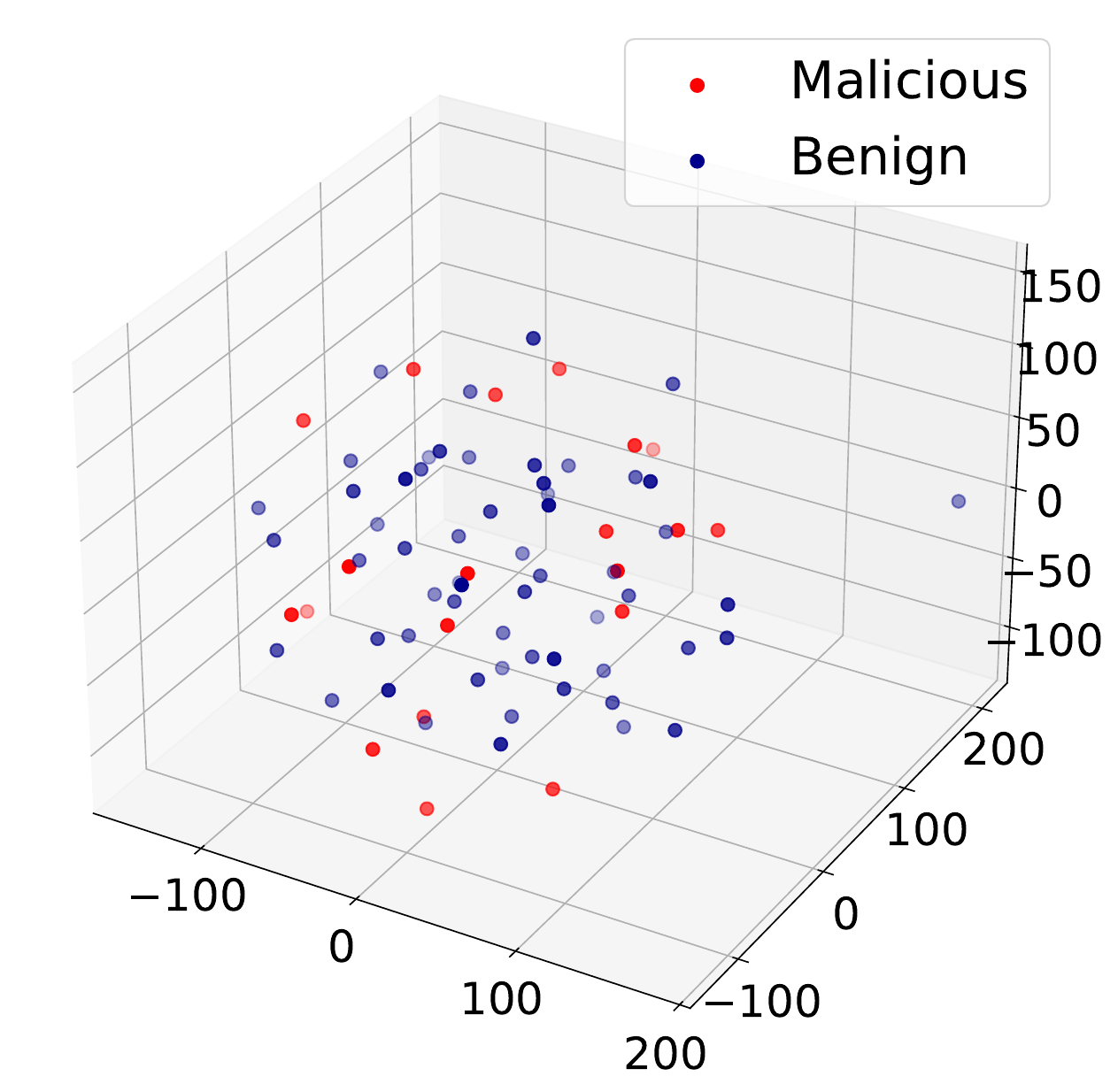}}
	\subfloat[Layer10]{\includegraphics[width=0.2\textwidth]{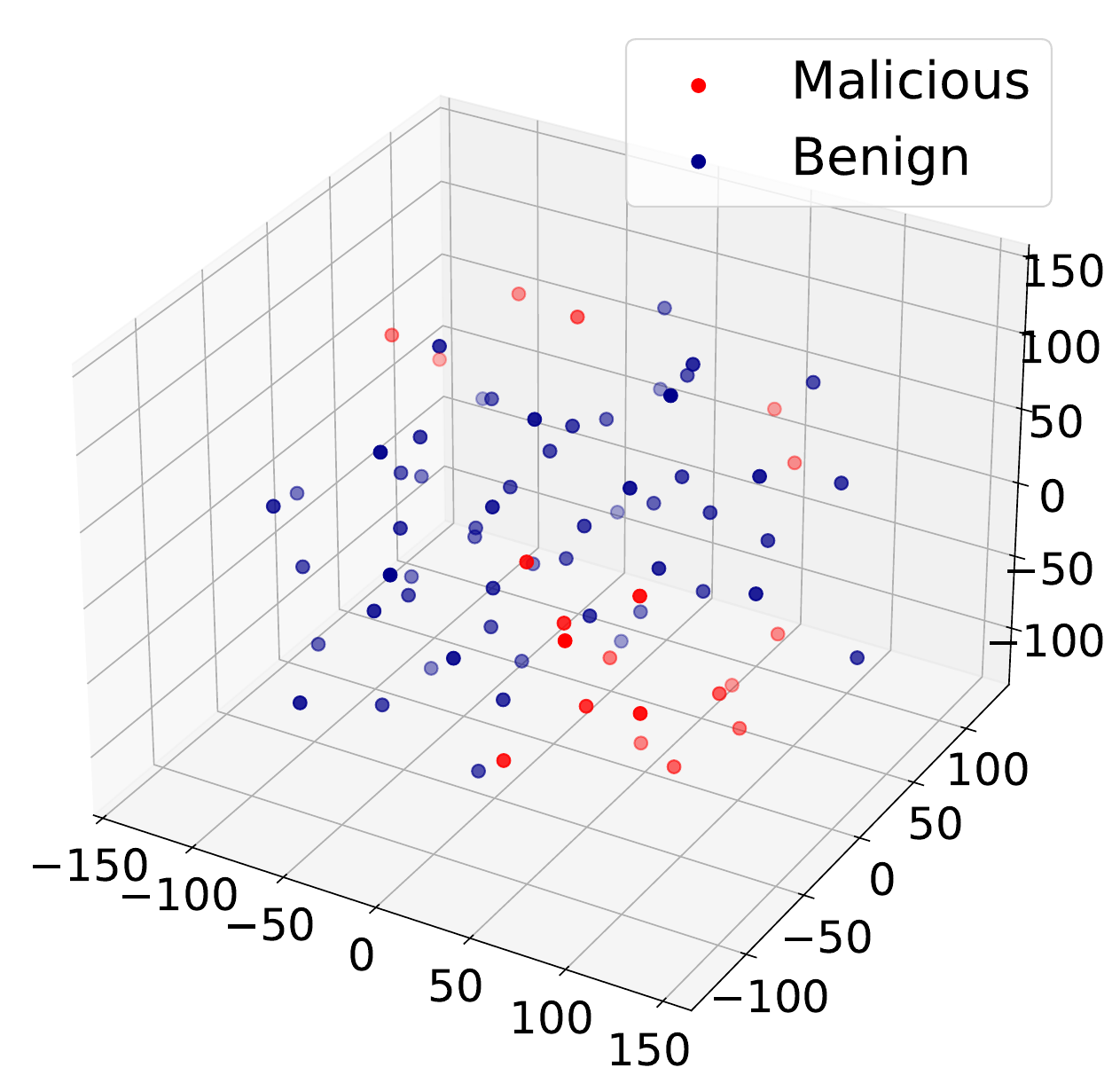}}\\
	\subfloat[Layer11]{\includegraphics[width=0.2\textwidth]{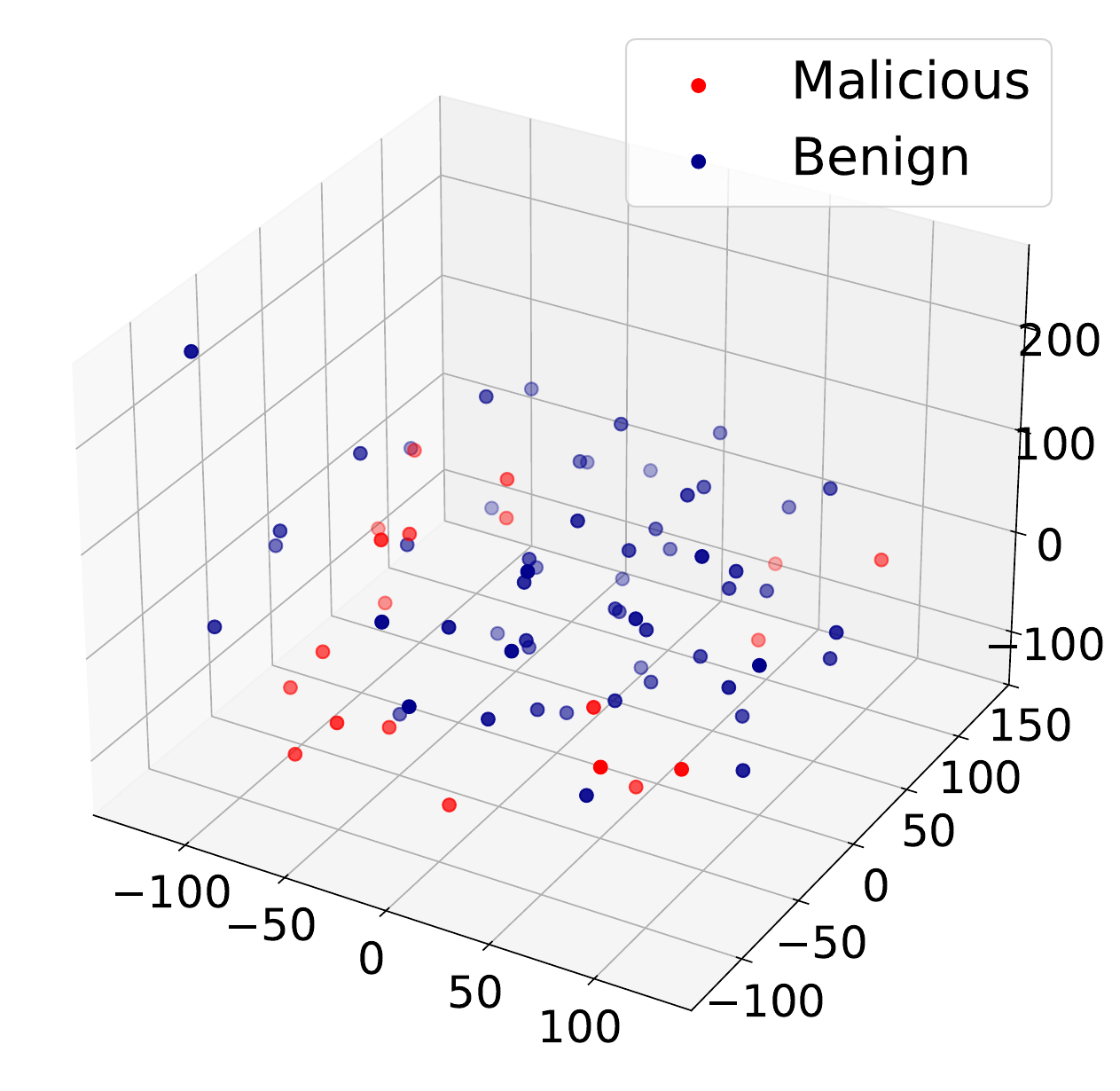}}
	\subfloat[Layer12]{\includegraphics[width=0.2\textwidth]{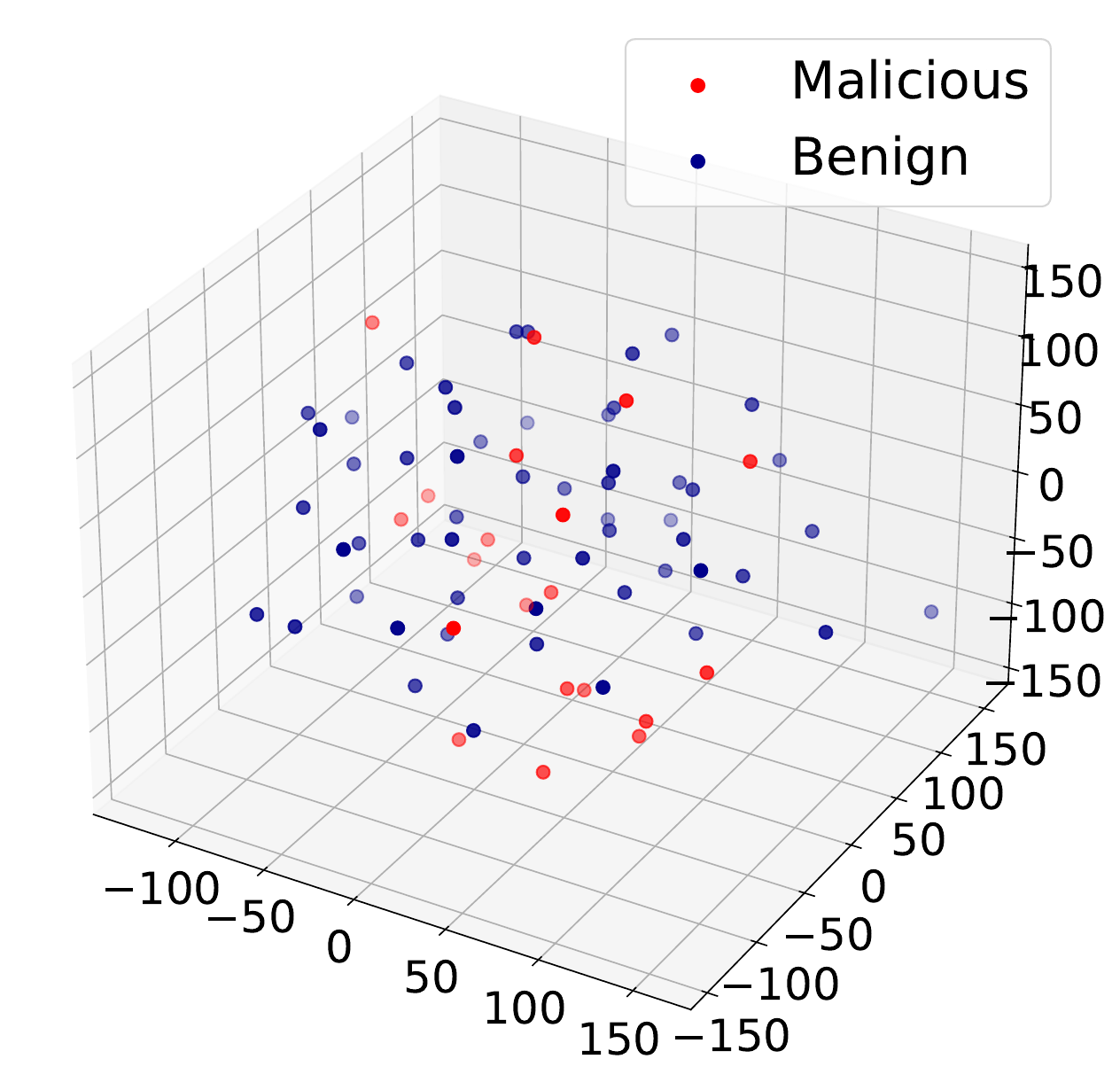}}
	\subfloat[Layer13]{\includegraphics[width=0.2\textwidth]{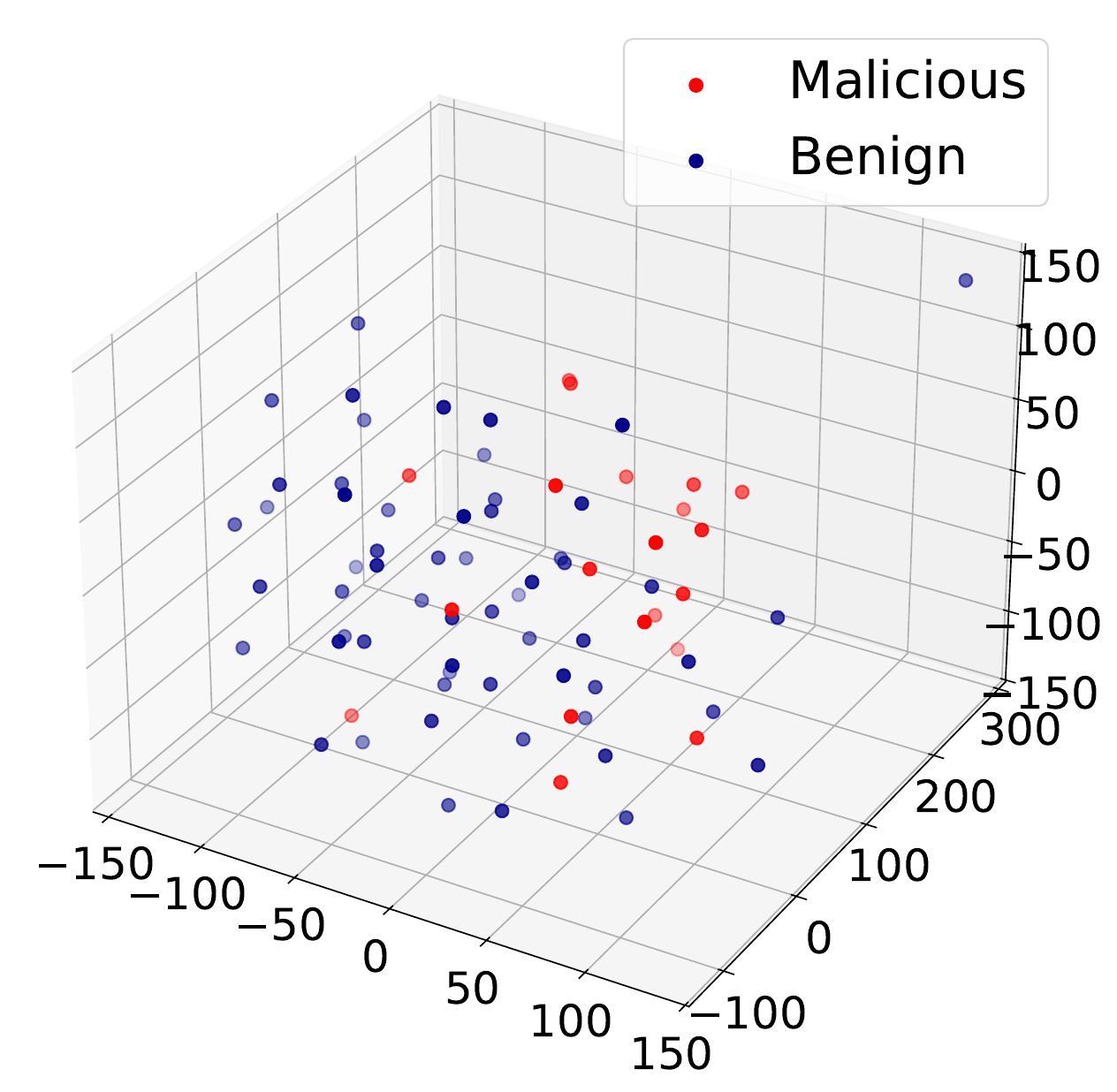}}
	\subfloat[Layer14]{\includegraphics[width=0.2\textwidth]{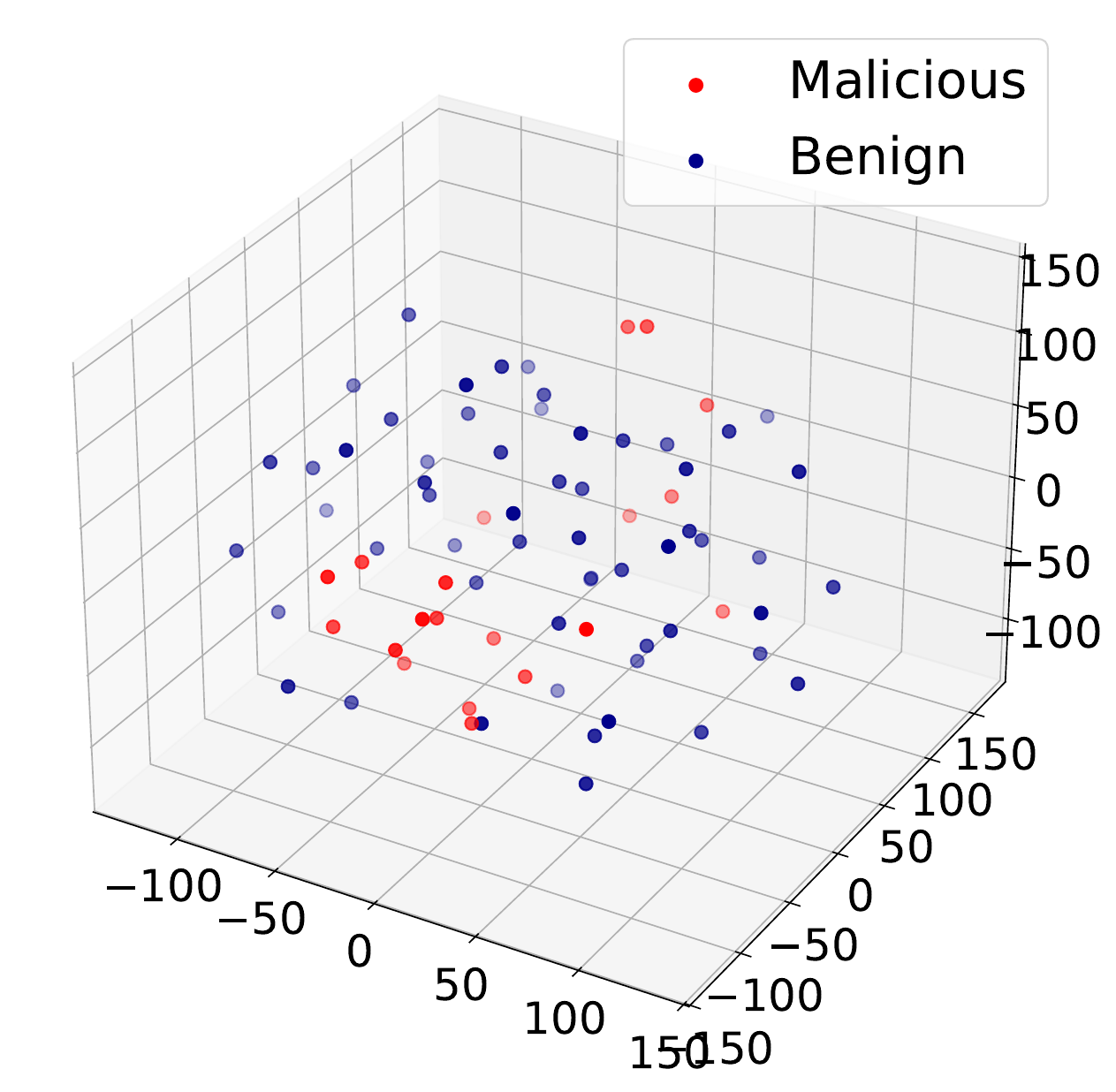}}
	\subfloat[Layer15]{\includegraphics[width=0.2\textwidth]{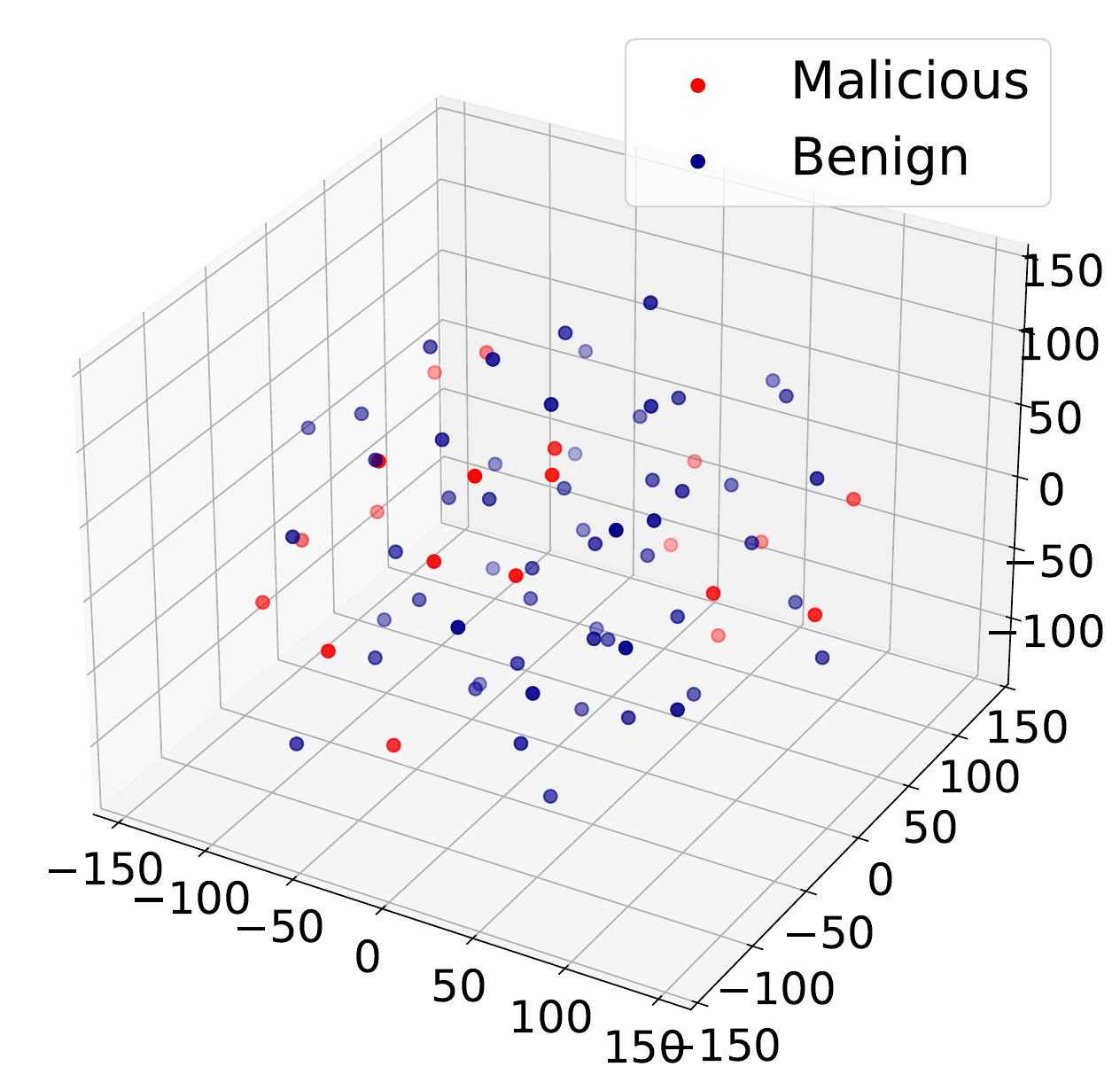}}\\
	\subfloat[Layer16]{\includegraphics[width=0.2\textwidth]{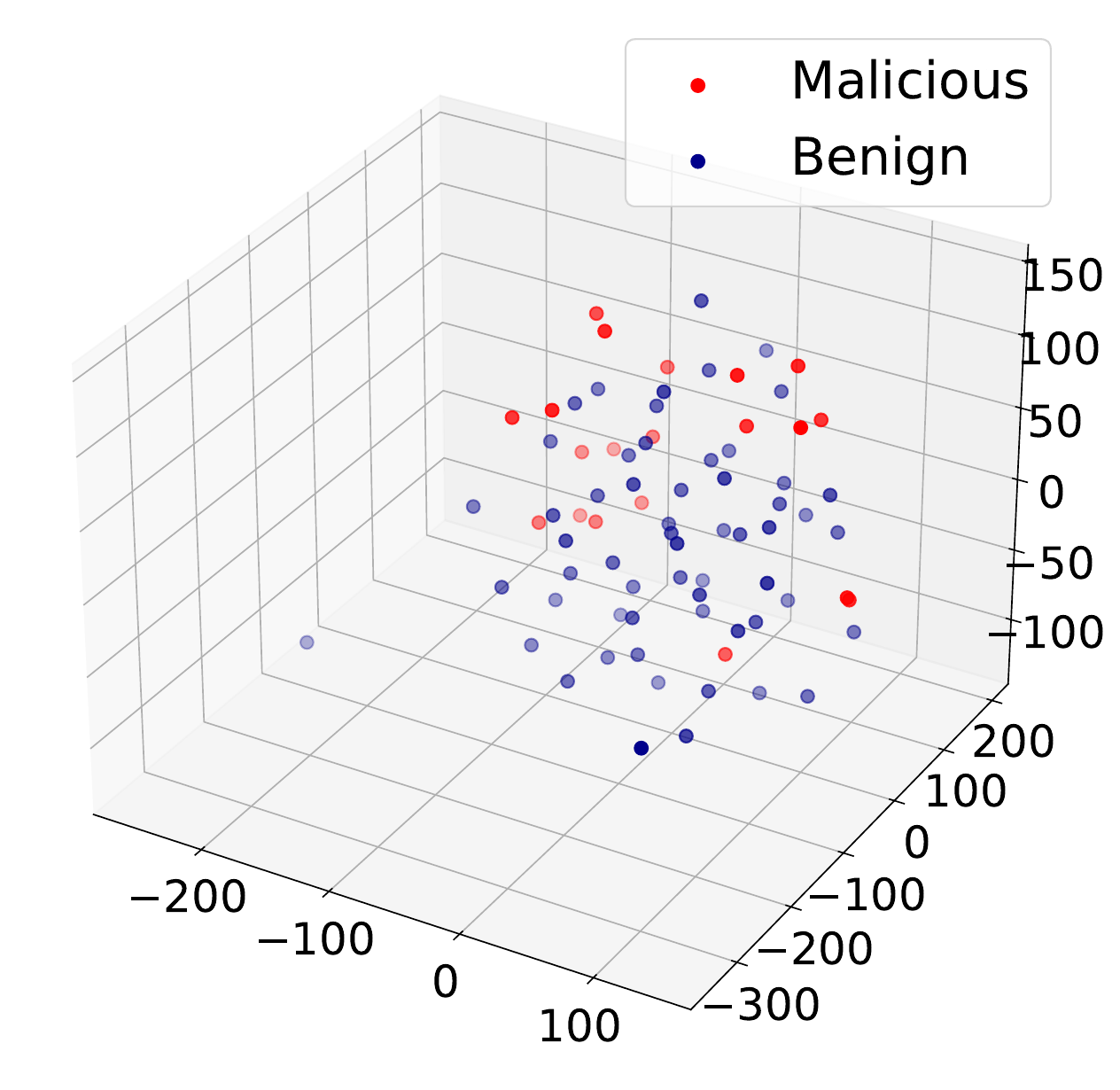}}
	\subfloat[Layer17]{\includegraphics[width=0.2\textwidth]{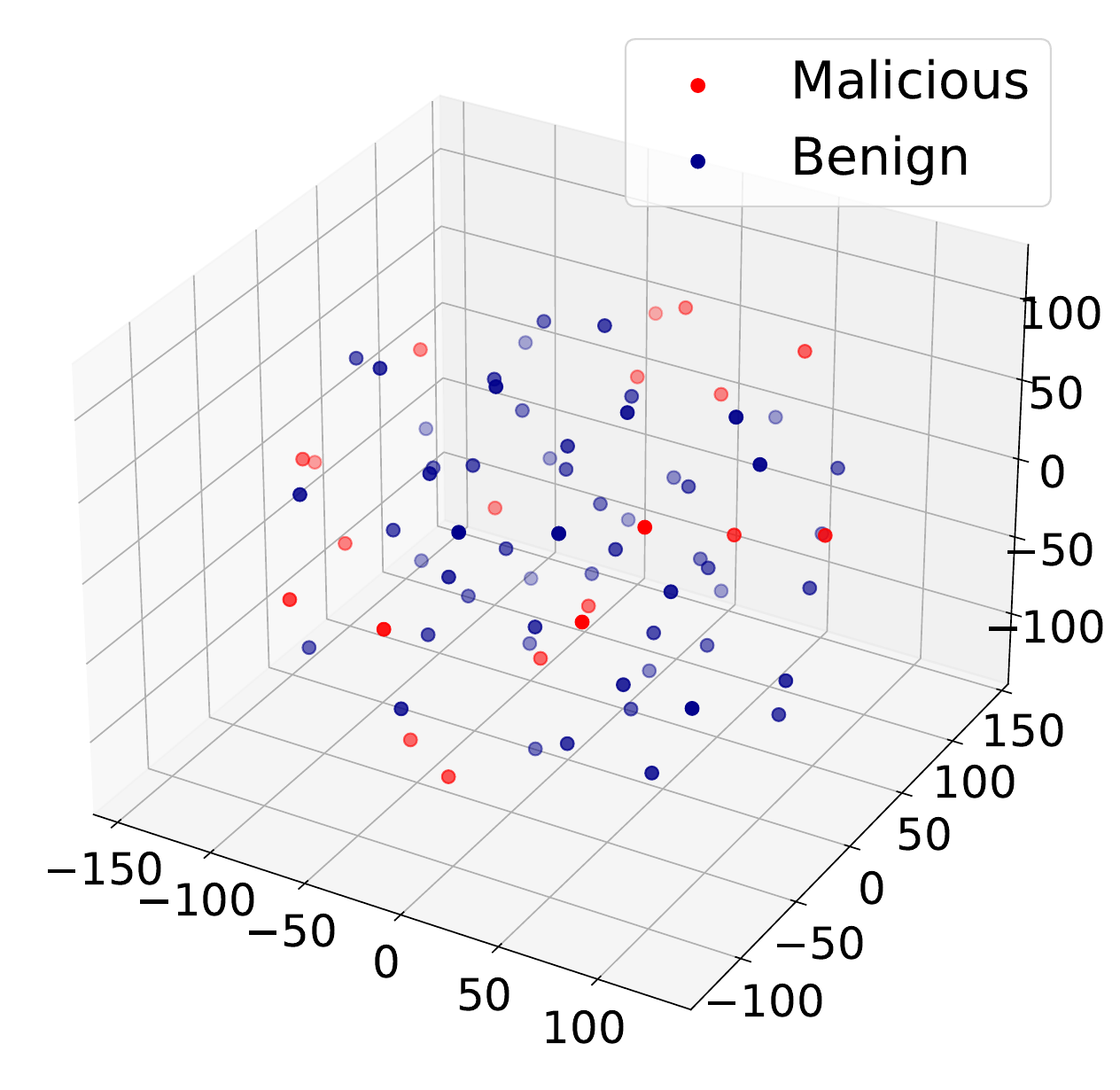}}
	\subfloat[Layer18]{\includegraphics[width=0.2\textwidth]{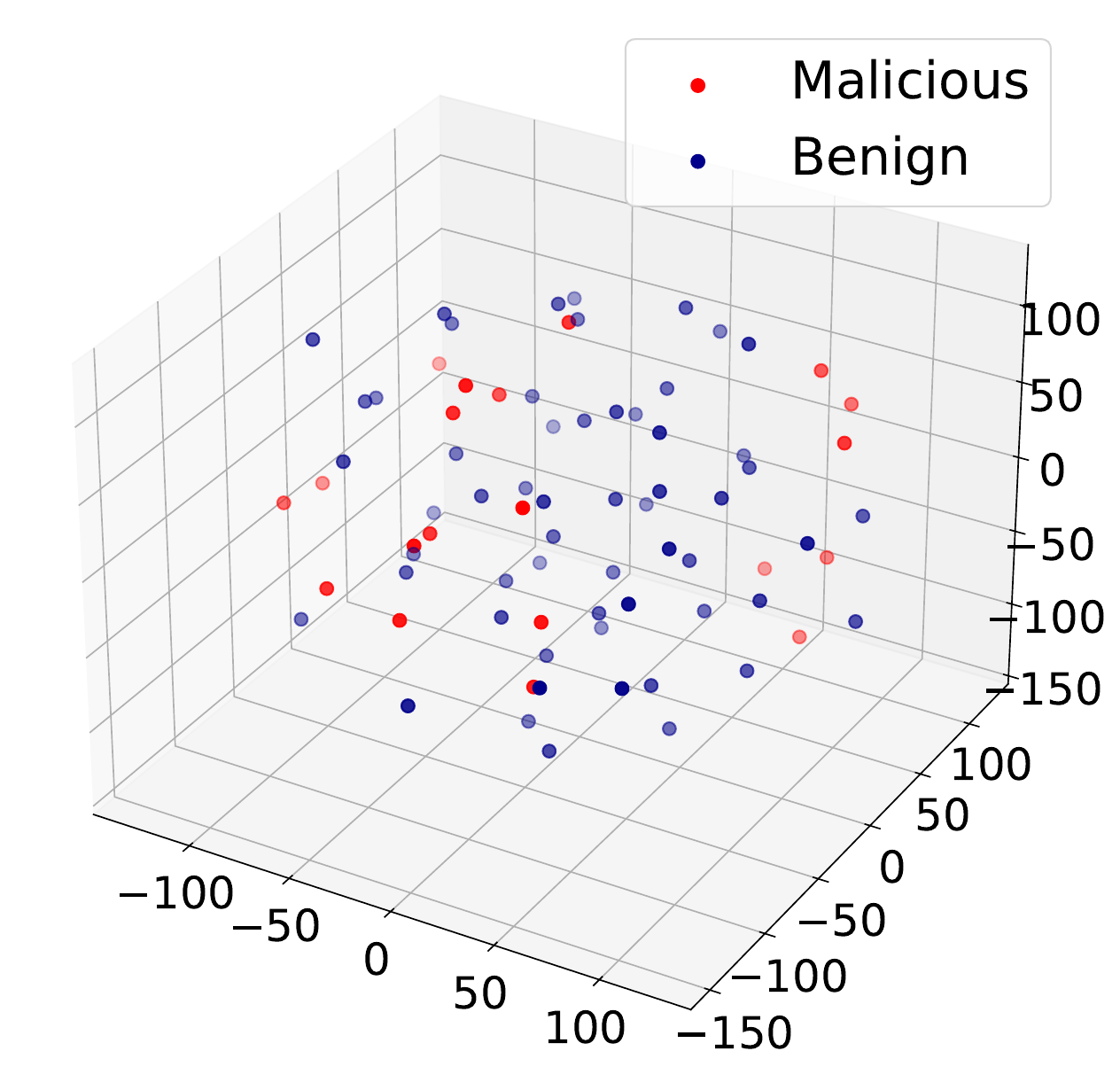}}
	\subfloat[Layer19]{\includegraphics[width=0.2\textwidth]{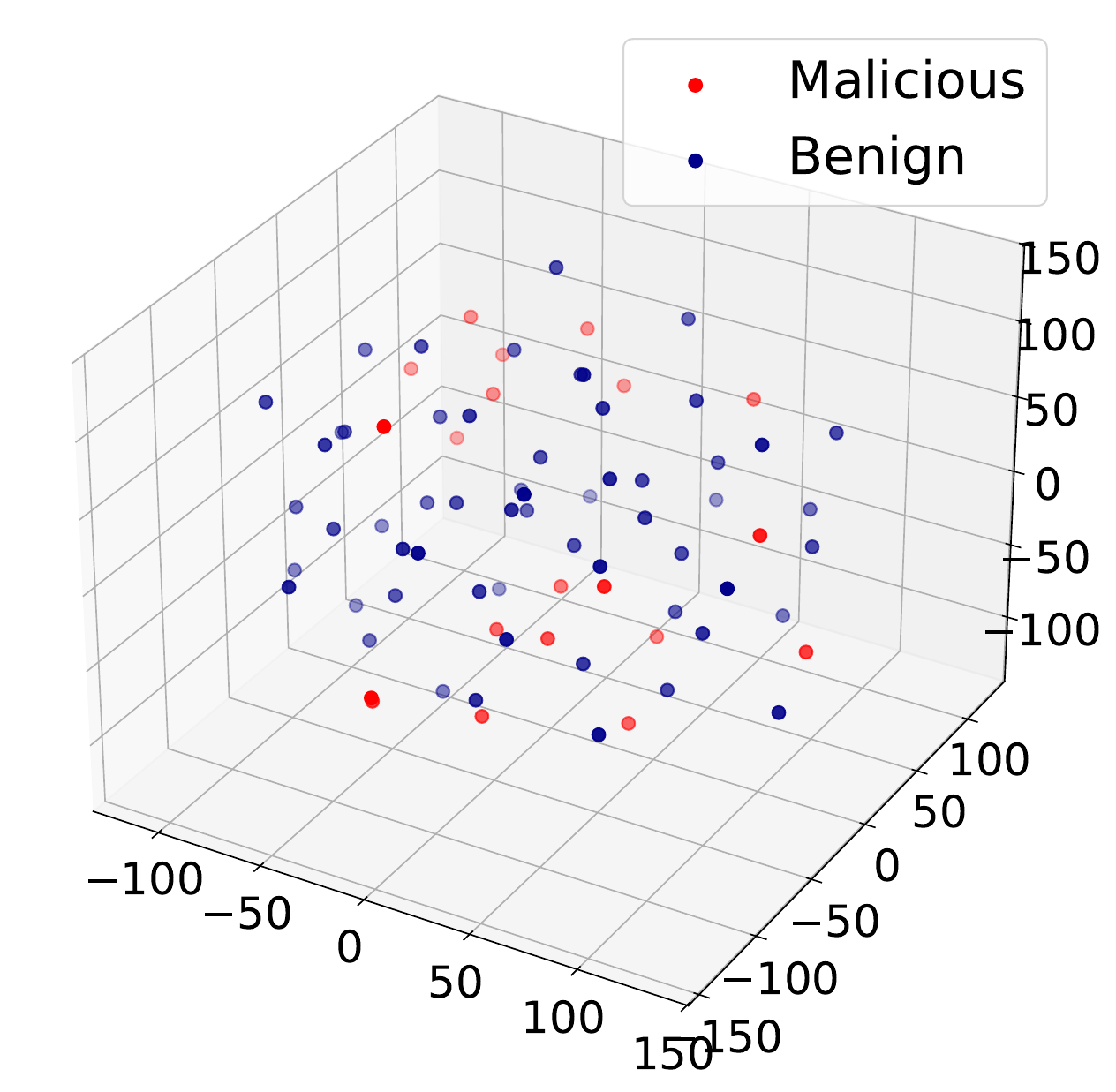}}
	\subfloat[Layer20]{\includegraphics[width=0.2\textwidth]{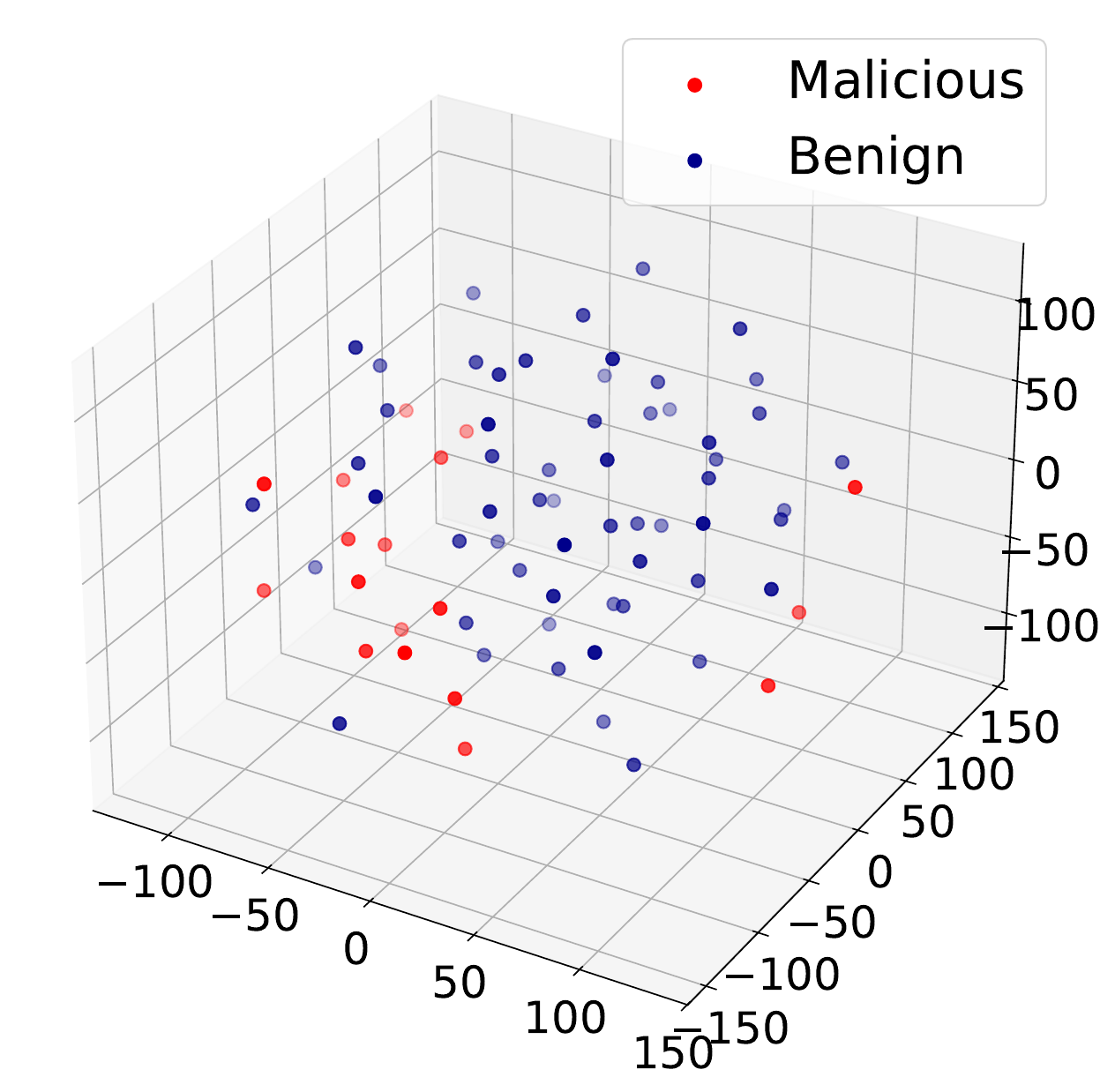}}\\
	\subfloat[Layer21]{\includegraphics[width=0.2\textwidth]{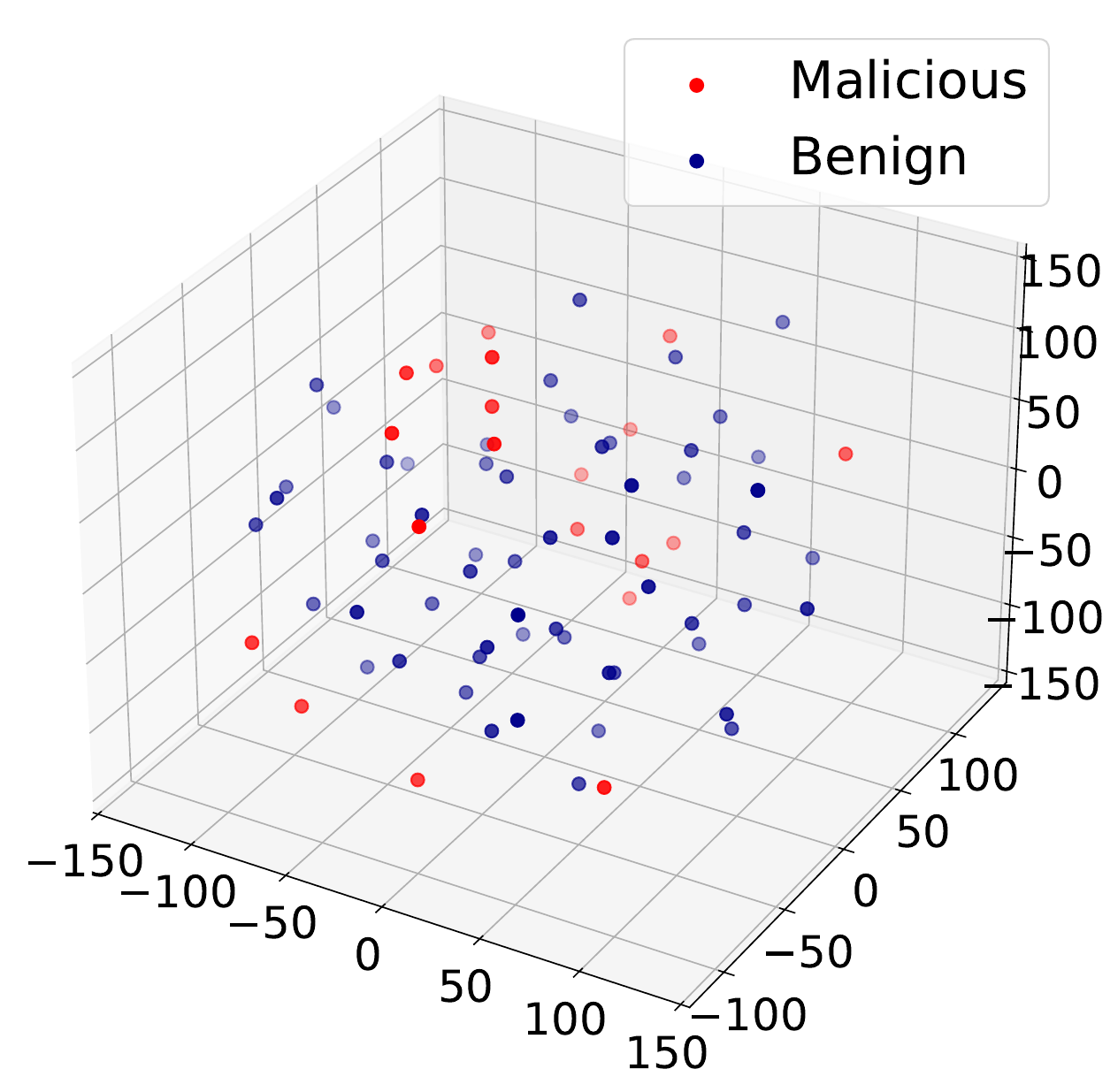}}
	\subfloat[Layer22]{\includegraphics[width=0.2\textwidth]{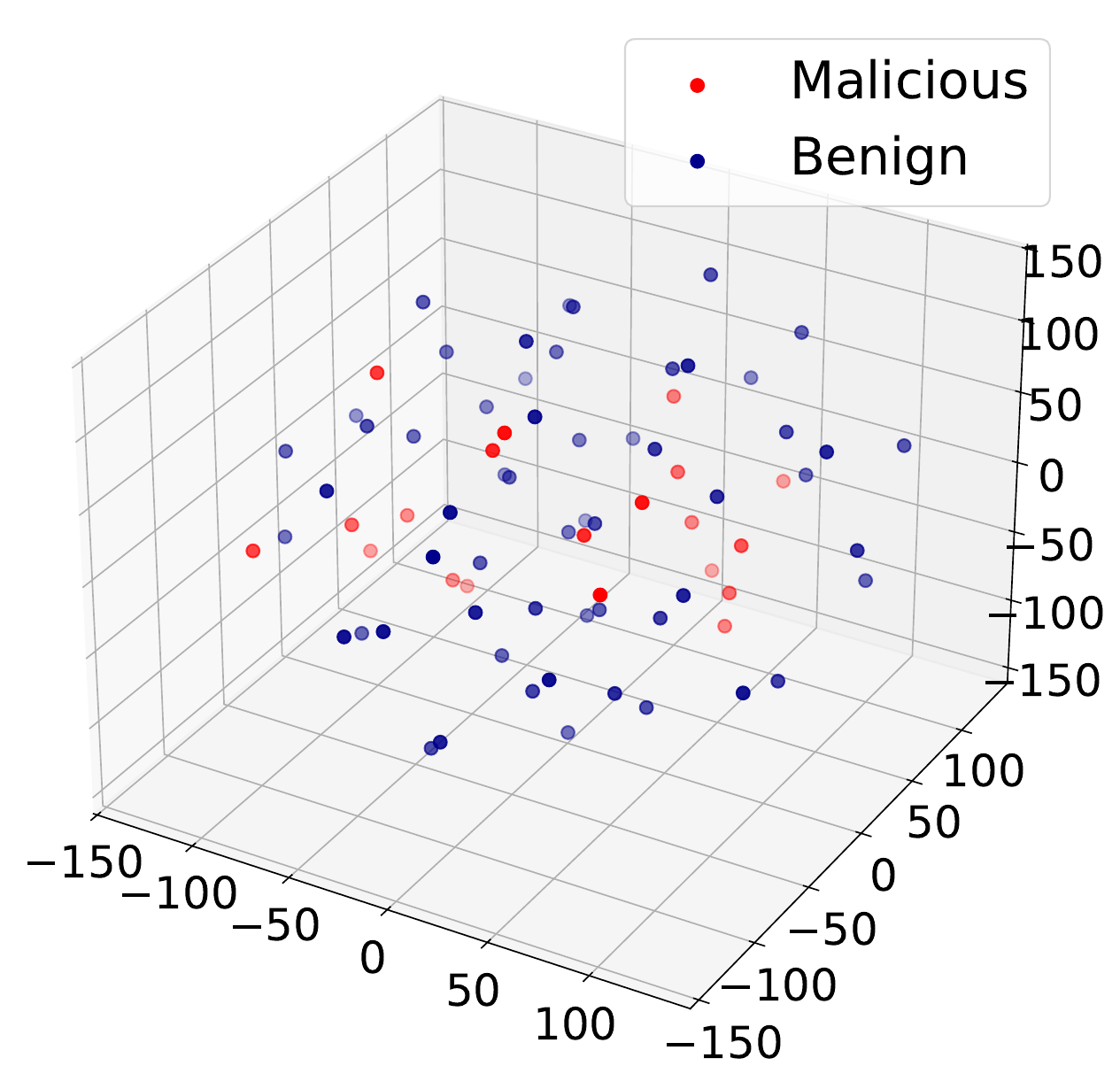}}
	\subfloat[Layer23]{\includegraphics[width=0.2\textwidth]{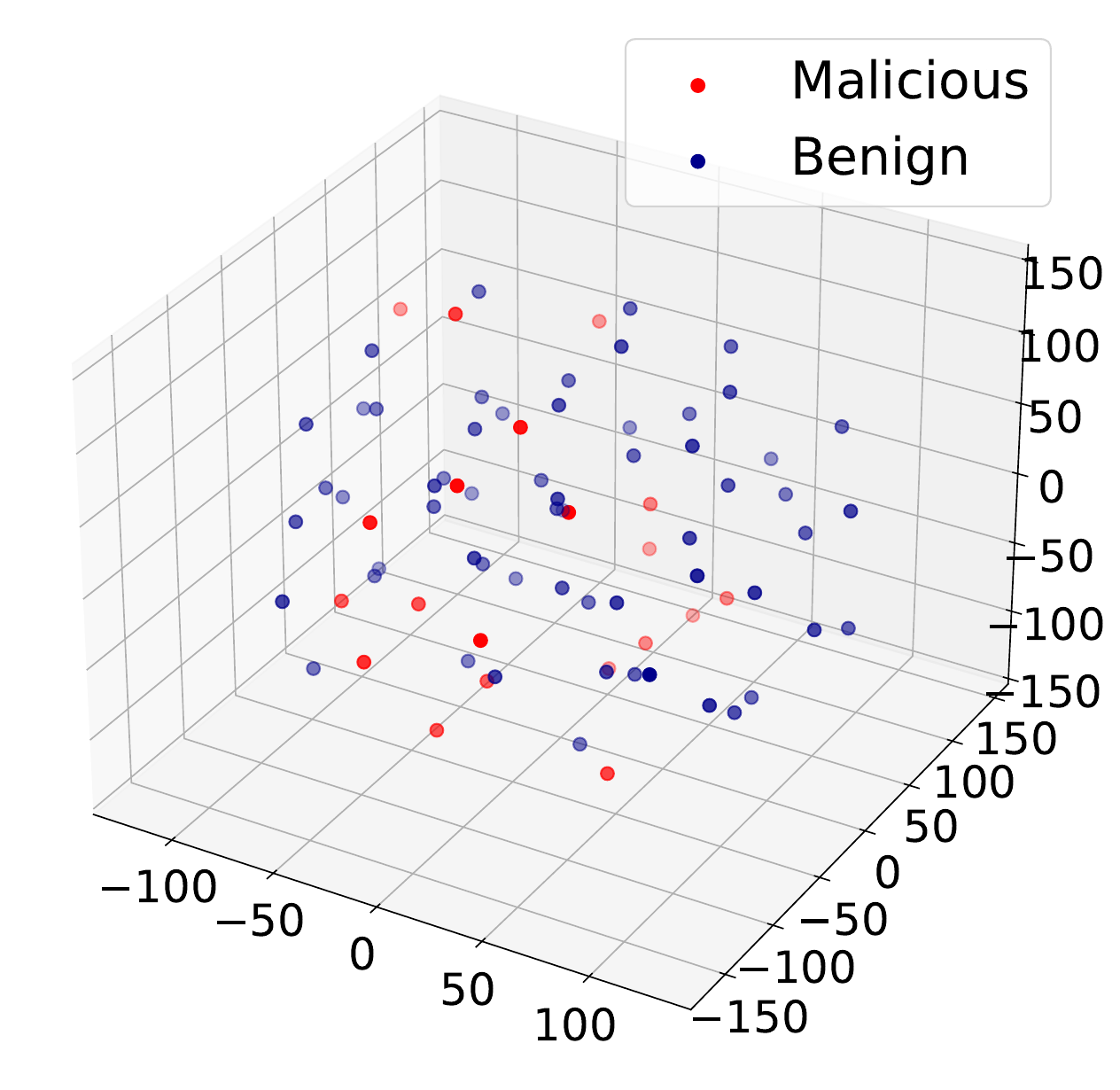}}
\\
	\caption{The layers of malicious models and benign models in the latent space.}
	\label{tsne}
\end{figure*}